\documentclass[12pt]{iopart}

\usepackage{iopams}
\usepackage{graphicx}
\usepackage{amssymb}
\usepackage{rotating}
\usepackage{pdflscape}
\usepackage{dcolumn}
\usepackage{bm}
\usepackage{cite}
\usepackage{longtable}
\usepackage[dvipsnames]{xcolor}
\begin{document}

\title[]{Decay properties of undetected superheavy nuclei with Z$>$110}
\author{A. Jain$^{1,2,3}$, P. K. Sharma$^{4}$, S. K. Jain$^{1}$, Dashty T. Akrawy$^{5,6}$, and G. Saxena$^{2,7,\dagger}$}
\address{$^{1}$Department of Physics, Manipal University Jaipur, Jaipur-303007, India}
\address{$^{2}$Department of Physics (H\&S), Govt. Women Engineering College, Ajmer-305002, India}
\address{$^{3}$Department of Physics, S. S. Jain Subodh P.G.(Autonomous) College, Jaipur-302004, India}

\address{$^{4}$Govt. Polytechnic College, Rajsamand-313324, India}
\address{$^{5}$Physics Department, College of Science, Salahaddin University, Erbil 44001-Kurdistan, Iraq}
\address{$^{6}$Becquerel Institute for Radiation Research and Measurements, Erbil, Kurdistan, Iraq}

\address{$^{7}$Department of Physics, Faculty of Science, University of Zagreb, Bijeni$\breve{c}$ka c. 32, 10000 Zagreb, Croatia.}

\ead{$^{\dagger}$gauravphy@gmail.com}
\vspace{10pt}
\begin{indented}
\item[]April 2023
\end{indented}

\begin{abstract}
A comprehensive study of favoured and unfavoured $\alpha$-decay, cluster decay, weak-decay along with spontaneous fission in undetected superheavy nuclei within the range for proton number 111$\leq$Z$\leq$118 and neutron number 161$\leq$N$\leq$192 is performed. Half-lives for various mentioned decays are estimated with good accuracy on the basis of NUBASE2020 and are found in excellent match with the known half-lives. $\alpha$-decay mode is found most probable in this wide range and correspondingly potential $\alpha$-decay chains are reckoned. Peculiarly, the chances of cluster emission, as well as weak-decay, are also anticipated in this region of the periodic chart which open new pathways of detection of superheavy nuclei.\par
\end{abstract}


\section{Introduction}
Nuclei with proton number Z$>$104, referred as superheavy nuclei (SHN), manifest the most exciting and challenging arena in the field of nuclear physics. Experimental facilities like GSI, Darmstadt \cite{hofmann2000,hofmann2011} and RIKEN, Japan \cite{morita2007} are available to synthesize SHN by cold-fusion reactions of $^{208}$Pb or $^{209}$Bi by beams of nuclei with mass number A$>$50 \cite{hamilton2013}. In the other experimental facility like Dubna laboratory, Oganessian \textit{et al} have successfully synthesized new SHN with Z=112$-$118 \cite{og2010,og2015npa} by the hot fusion reaction with $^{48}$Ca beam and various actinide targets. The last synthesized element up to so far has proton number Z$=$118 \cite{Og2006}, nevertheless, many of the SHN are still undetected in the laboratory for which several experimental attempts have already been aimed \cite{dullmann2014,heinz2012,Og2009,hofmann2016}. In this connection, recently, few experiments have been performed for the future possibilities of synthesizing new elements with Z$=$119 and Z$=$120 \cite{voinov2020}.\par

The SHN are highly unstable and can break via various decays viz. $\alpha$-decay, spontaneous fission, cluster decay, weak-decay, etc. Among these decay modes, $\alpha$-decay plays a crucial role in the identification of SHN in the laboratories through $\alpha$-decay chains \cite{CHENG2019,hofmann1984}. To plan experiments related to the identification of SHN, various theoretical inputs are required out of which
estimation of $\alpha$-decay half-lives is most pivotal. Various theoretical methods and models have been employed to estimate $\alpha$-decay half-lives such as Gamow-like model (GLM) \cite{Zdeb2013}, fission-like model \cite{yong2010}, liquid drop model \cite{poenaru1979,royer2000} with its modifications \cite{cui2018,bao2014,royer2008,santhosh2020}, and Coulomb and proximity potential model (CPPM) \cite{zanganah2020NPA,santhosh2012NPA}.\par

The half-lives for $\alpha$-transition can also be calculated from various empirical formulas based on Geiger-Nuttall law. There are several empirical or semi-empirical formulas \cite{vss1966,sobi1989,parkho2005,brown1992,renA2004,qi2009,saxena2021} along with their refitted or modified versions \cite{xu2022,akrawy2017,akrawy2019,singh2020,Saxena2021jpg,akrawy2022EPJA} which are widely being used to estimate $\alpha$-decay half-lives in different regions of the periodic chart. In the present study, we have estimated $\alpha$-decay half-lives of 146 (even-even and odd-A) undetected SHN within the range 111$\leq$Z$\leq$118 and 161$\leq$N$\leq$192 by using one of the recent modified empirical formula, i.e. new modified Horoi formula (NMHF) \cite{sharma2021npa}, after testing its accuracy on the 424 experimental half-lives corresponding to ground to ground favoured and unfavoured $\alpha$-transitions. We have also compared our theoretical half-lives with that of the experimental half-lives of the known $\alpha$-decay chains \cite{og2017,og1999,og2015npa,forsberg2016,og2000,audi20201} in the above mentioned range. These estimates from NMHF are found in excellent match and therefore the formula is utilized to predict half-lives of the future potential SHN related to the already known or undetected decay chains.\par

Another decay mode, i.e. spontaneous fission (SF) which was firstly discussed by Bohr and Wheeler \cite{Bohr1939} and experimentally verified by Flerov and Petrjak \cite{flerov1940}, is also found to be equally decisive in Z$\geq$90 nuclei \cite{dvorak2006,og2005}. The first theoretical try to estimate the half-lives of SF was done by Swiatecki in 1955 with liquid drop model \cite{swiatecki1955} and uptill now many empirical formulas are proposed to compute half-life of SF \cite{ren2005,xu2008,bao2015,santhosh2016,soylu2019CPC,santhosh2021SF,Hessberger2017,Reinhard2018}. Recently, the Bao formula \cite{bao2015} has been modified (modified Bao formula (MBF) \cite{Saxena2021jpg}) using the latest evaluated nuclear properties table NUBASE2020 \cite{audi20201} and is used in the present work for the calculation of half-lives of SF.\par

The possibility of the heavy particle radioactivity (cluster decay) was also predicted in the heavy and superheavy nuclei \cite{poenaru2012,santhosh2021,santhosh2019,soylu2021,sowmya2021}. Discussion of cluster decay started by Sandulescu \textit{et al} in 1980 \cite{sandulescu1980} and firstly observed by Rose and Jones \cite{rose1984}. Cluster decay refers to the decay of a fragment from the parent nucleus. Decay of fragments like $^{20}$O, $^{22,24-26}$Ne, $^{28-30}$Mg, and $^{32,34}$Si \cite{barwick1986,bonetti2007} have been already observed experimentally for the nuclei with Z ranging from 87 to 96. The decay of such clusters is firmly connected to closed shell daughters as $^{208}$Pb or its neighbours, and therefore predominantly diverge towards Pb isotopes. In superheavy nuclei, the probability of heavy clusters viz. Se, Br, Kr, Rb, Sr, Y, Zr, Nb, Mo etc. is already speculated in several Refs. \cite{zhang2018,soylu2019,santhosh2018,poenaru2018,poenaru2018EPJA,matheson2018,Warda2018,jain2021}. To invoke the competition of various decay modes, we have comprehensively investigated cluster decay half-lives by using universal decay law (UDL) \cite{udl2009} for all the possible isotopes between
111$\leq$Z$\leq$118 and 161$\leq$N$\leq$192. The clusters considered for these nuclei are chosen to result Pb isotopes (Z$=$82) as possible daughter nuclei which lead to the emitting clusters ranging from Z$=$29 to Z$=$36.\par

One more decay mode is the weak-decay ($\beta^{-}$/$\beta^{+}$/EC-decay) \cite{og2011,karpov2012,zhang2006,zhang2007,Sobhani,frdm2019} which is a principal decay mode in a broad region of the periodic chart, however, is rarely found to exist in the superheavy region. Despite this fact, the possibility of weak-decay is also contemplated in the superheavy region by several Refs. \cite{hirsch1993,singh2020,sarriguren2019,sarriguren2018}. In the present study of SHN, we have also examined the weak-decay that is found somewhat apparent in few of the SHN consist of the isotopes of Rg, Cn, Nh, and Fl. The half-lives of weak-decay are calculated by using an empirical formula given by Fiset and Nix \cite{Fiset1972} and recently proven to be simple and successful in several SHN \cite{saxenaijmpe2019,singh2020,sharma2022}.\par

Finally, we have calculated half-lives and the competition of all probable decay modes considering them on equal footing, by using optimally chosen respective empirical formulas. The sensitivity of these half-lives on the uncertainties of $Q$-values are also evaluated by using available experimental uncertainties \cite{og2017,og1999,og2015npa,forsberg2016,og2000,audi20202}. Since, the present work is focussed on superheavy nuclei where the role of uncertainties in $Q$-values becomes very crucial, therefore, we have evaluated theoretical uncertainties in $Q$-values of $\alpha$-decay taken from WS4 mass model \cite{ws42014} by using 69 experimental data \cite{audi20202} for Z$\geq$106. Due to unavailability or sufficient experimental data of $Q$-values for $\beta^{-}$-decay, EC decay, and cluster decay, we have calculated the respective uncertainties in theoretical $Q$-values by using 113, 78, and 73 experimental data \cite{barwick1986,bonetti2007,audi20202} for Z$>$82. The results of half-lives of each decay are found in an excellent match with the available half-lives and hence utilised to estimate the half-lives of several unknown (undetected) nuclei within the range 111$\leq$Z$\leq$118 and 161$\leq$N$\leq$192. This kind of contest of several decay modes after their accurate estimation of half-lives is an intriguing platform for the experiments eyeing up for the detection of new SHN.
\section{Formalism}
\subsection{$\alpha$-decay}
For the selection of $\alpha$-decay half-lives formula, we have tested few recently fitted/modified empirical/semiempirical formulas
\cite{sharma2021npa,saxena2021,newrenA2019,IRF2022,MYQZR2019,royer2020,akrawy2019,soylu2021,akrawy2018,akrawy2018ijmpe,Akrawymrf2018,akrawy2022EPJA,UF2022,Ismail2022,cheng2022}
and the comparison among these formulas is shown in Table \ref{RMSE}. For the comparison, we use root mean square error (RMSE) and uncertainty ($u$), which are calculated by using the following formulas:
\begin{eqnarray}
RMSE &=& \sqrt{\frac{1}{N_{nucl}}\sum^{N_{nucl}}_{i=1}\left(logT^i_{th}-logT^i_{exp}\right)^2}\\
\label{RMSE}
u &=& \sqrt{\frac{\sum(x_{i}-\upsilon)^2}{N_{nucl}(N_{nucl}-1)}}
\label{uncer}
\end{eqnarray}
here, $N_{nucl}$ is the total number of data. $T^i_{th}$ and $T^i_{exp}$ are the theoretical and experimental values of half-life for $i^{th}$ data point, respectively. $x_{i}$ is the $i^{th}$ reading of data set, $\upsilon$ is the mean of the data set.
We found that NMHF has minimum RMSE (0.56) and uncertainties ($\pm$0.09 s) as tested it on 40 experimental data within the range 111$\leq$Z$\leq$118 and 161$\leq$N$\leq$192 while compared to other similar, latest fitted and known formulas such as Royer2020 \cite{royer2020}, MYQZR2019 \cite{MYQZR2019}, Akrawy2018 \cite{akrawy2018}, Modified RenB2019 \cite{newrenA2019}, MRF2018 \cite{Akrawymrf2018}, DK2018 \cite{Akrawymrf2018}, Modified Budaca Formula (MBuF2022) \cite{akrawy2022EPJA}, UF2022 \cite{UF2022}, IUF2022 \cite{Ismail2022}, ISEF2022 \cite{cheng2022}, new modified Sobiczewski formula (NMSF2021) \cite{sharma2021npa} and new modified Manjunatha formula (NMMF2021) \cite{sharma2021npa}.\par

\begin{table}[!htbp]
 \caption{Comparison of RMSE and uncertainties (in second) for the different tested formulae on 40 experimental data within the range 111$\leq$Z$\leq$118 and 161$\leq$N$\leq$192.}
 \centering
  \resizebox{1.0\textwidth}{!}{%
 \begin{tabular}{@{\hskip 0.4in}c@{\hskip 0.4in}c@{\hskip 0.4in}c|@{\hskip 0.4in}c@{\hskip 0.4in}c@{\hskip 0.4in}c}
 \hline
 \hline
 Formula                            &  RMSE   & Uncertainty  &  Formula                            &   RMSE & Uncertainty  \\
   \hline
NMHF2021 \cite{sharma2021npa}       &   0.56  &  $\pm$0.09        & MUDL2019 \cite{akrawy2019}          &   1.03 & $\pm$0.16 \\
QF2021 \cite{saxena2021}            &   0.70  &  $\pm$0.11        & NMSF2021 \cite{sharma2021npa}       &   1.12 & $\pm$0.12 \\
IUF2022 \cite{Ismail2022}           & 0.76    &  $\pm$0.11        & MYQZR2019 \cite{MYQZR2019}          &   1.20 & $\pm$0.15 \\
MTNF2021 \cite{saxena2021}          &   0.80  &  $\pm$0.12        & Soylu2021 \cite{soylu2021}          &   1.25 & $\pm$0.13 \\
NRenA2019 \cite{newrenA2019}        &   0.81  &  $\pm$0.12        & DK2018 \cite{akrawy2018}            &   1.33 & $\pm$0.16 \\
IRF2022 \cite{IRF2022}              &   0.84  &  $\pm$0.12        & MBuF2022 \cite{akrawy2022EPJA}      &   1.41 & $\pm$0.17 \\
ISEF2022 \cite{cheng2022}           &   0.95  &  $\pm$0.13        & NMMF2021 \cite{sharma2021npa}       &   1.43 & $\pm$0.15 \\
Royer2020 \cite{royer2020}          &   0.95  &  $\pm$0.14        & Akrawy2018 \cite{akrawy2018ijmpe}   &   1.56 & $\pm$0.19 \\
MSLB2019 \cite{MYQZR2019}           &   0.95  &  $\pm$0.14        & MRenB2019 \cite{newrenA2019}        &   1.81 & $\pm$0.23 \\
UF2022 \cite{UF2022}                &   1.00  &  $\pm$0.15        & MRF2018 \cite{Akrawymrf2018}        &   1.88 & $\pm$0.21 \\

 \hline
\hline \end{tabular}}
\label{RMSE}
\end{table}

Consequently, the present calculations of $\alpha$-decay half-lives are performed by using NMHF \cite{sharma2021npa} which incorporates the terms related to (i) angular momentum taken away by the $\alpha$-particle and (ii) asymmetry of the parent nucleus. This formula is found to produce precise $\alpha$-decay half-lives for full periodic chart ranging from heavy to superheavy nuclei \cite{sharma2021npa}. The formula is given by:
\begin{eqnarray}
log_{10}T_{1/2}^{NMHF}(s) &=& (a\sqrt{\mu} + b)[(Z_{\alpha}Z_{d})^{0.6}Q_{\alpha}^{-1/2} - 7] + (c\sqrt{\mu} + d) + eI + fI^{2}\nonumber\\ && + gl(l+1)
\label{qf}
\end{eqnarray}
where the coefficients a, b, c, d, e, f, and g obtained by fitting are 107.0131, -206.5398, -160.6152, 309.6165, 19.7237, -31.1655, and 0.0238, respectively. Also, $Z_{d}$ represents atomic number of daughter nucleus, $\mu$ is the reduced mass which is given by $A_{d}A_{\alpha}/(A_{d}+A_{\alpha})$ where $A_{d}$ and $A_{\alpha}$ are the mass number of daughter nucleus and $\alpha$-particle, and $I$ (=(N-Z)/A) is the nuclear isospin asymmetry. For this work, if the $Q_{\alpha}$ values are not available in atomic mass evaluation tables AME2020 \cite{audi20202} then the theoretical $Q_{\alpha}$ values from WS4 mass model \cite{ws42014} are used which are found quite accurate in comparison to few other mass models \cite{saxena2021}. The minimum angular momentum $l$ of $\alpha$-particle, which distinguishes between favoured and unfavoured $\alpha$-transitions, can be obtained by standard selection rules \cite{denisov2009} using spin and parity values of the parent and daughter nuclei.
 \begin{eqnarray}
   l=\left\{
    \begin{array}{ll}
       \triangle_j\,\,\,\,\,
       &\mbox{for even}\,\,\triangle_j\,\mbox{and}\,\,\pi_{p} = \pi_{d}\\
       \triangle_{j}+1\,\,\,\,\,
       &\mbox{for even}\,\,\triangle_j\,\mbox{and}\,\,\pi_{p} \neq \pi_{d}\\
       \triangle_{j}\,\,\,\,\,
       &\mbox{for odd}\,\,\triangle_j\,\mbox{and}\,\,\pi_{p} \neq \pi_{d}\\
       \triangle_{j}+1\,\,\,\,\,
       &\mbox{for odd}\,\,\triangle_j\,\mbox{and}\,\,\pi_{p} = \pi_{d}\\
      \end{array}\right.
      \label{selection-rules}
\end{eqnarray}
where $\triangle_j$=$|j_p - j_d|$ with j$_{p}$, $\pi_{p}$, j$_{d}$, $\pi_{d}$, being the spin and parity values of the parent and daughter nuclei, respectively. For the present paper, spin and parities are taken from Ref. \cite{audi20201}, if available, otherwise taken theoretically from Ref. \cite{moller2019}.\par

\subsection{Spontaneous fission (SF)}
In this part of the periodic chart, spontaneous fission (SF) is also found as probable as that of $\alpha$-decay, which eventually ascertained crucial for the planning of experiments towards the detection of new elements/isotopes through $\alpha$-decay chains. Therefore, the contest with SF has also accounted by testing several available empirical formulas on the experimentally known half-lives \cite{audi20201} of SF for the nuclei Z$>$110. There are only 5 nuclei ($^{281}$Rg, $^{282,284}$Cn, and $^{284,286}$Fl) with known experimental half-lives of their SF and are used for probing the accuracy of RenA formula \cite{ren2005}, formula by Xu \textit{et al} \cite{xu2008}, modified Swiatecki formula \cite{bao2015}, semi-empirical formula proposed by Santhosh \textit{et al} \cite{santhosh2016}, Soylu formula \cite{soylu2019CPC}, modified Santhosh formula \cite{santhosh2021SF} and Modified Bao formula (MBF) \cite{Saxena2021jpg}. It is found that among the mentioned formulas, MBF estimates the half-lives more accurately as the RMSE values for available experimental 35 data (all) and 5 data (for the range 111$\leq$Z$\leq$118) \cite{audi20202} are found minimum i.e. 1.42 and 1.86, respectively. Therefore, the MBF formula is used in the present work which is expressed as below \cite{Saxena2021jpg}:
\begin{eqnarray}
log_{10}T_{1/2}^{SF} (s) &=& c_1 + c_2 \left(\frac{Z^2}{(1-kI^2)A}\right)+ c_3\left(\frac{Z^2}{(1-kI^2)A}\right)^2 + c_4 E_{s+p}
\label{baoSF}
\end{eqnarray}

here k$=$2.6 and other coefficients are c$_2$$=$$-$37.0510, c$_3$$=$0.3740, c$_4$$=$3.1105. The values of c$_1$ for various sets of nuclei are
c$_1$(e-e)$=$893.2645, c$_1$(e-o)$=$895.4154, c$_1$(o-e)$=$896.8447 and c$_1$(o-o)$=$897.0194.

\subsection{Cluster decay}
To calculate the logarithmic half-lives of cluster emission, there are various available empirical formulas viz. Royer formula \cite{Royercluster}, the formula given by Balasubramaniam \textit{et al} (BKAG) \cite{BKAG}, Horoi formula \cite{horoi2004}, the formula by Ren \textit{et al} \cite{renA2004}, NRDX formula \cite{nrdx2008}, universal decay law (UDL) \cite{udl2009}, universal curve formula (UNIV) \cite{UNIV}, Tavares-Medeiros formula (TM) \cite{TM}, formula by Soylu \cite{Soylu2021}, improved unified formula (IUF) \cite{Ismail2022}, improved semi-empirical formula (ISEF) \cite{cheng2022} and some recent modified formulas \cite{Jain2023}. Considering the absence of experimental data of cluster radioactivity in superheavy region, only UDL formula among the above mentioned formulas demonstrates for the existence of a competitive relationship between $\alpha$-decay and cluster radioactivity in superheavy region, due to its treatment as both the
preformation model and the fission-like mechanisms \cite{Zhang2018}, and henceforth justifies its use for further predictions. The UDL formula is represented by:
\begin{eqnarray}
log_{10}T_{1/2}^{UDL}(s) &=& aZ_{c}Z_{d}\sqrt{\frac{\mu}{Q}}+b[\mu Z_{c}Z_{d}({A_{c}}^{1/3}+
{A_{d}}^{1/3})]^{1/2}+c
\label{udl}
\end{eqnarray}
where a, b and c are the fitting coefficients and the values of these coefficients are 0.3949, -0.3693 and -23.7615, respectively. $\mu$ is reduced mass and calculated by the formula given as $A_{d}A_{c}/(A_{d}+A_{c})$ where d and c subscripts represent quantities related to daughter nucleus and emitted cluster, respectively. \par

\subsection{Weak-decay}
To look into the possibility of weak-decay, we have picked up experimental half-lives of $\beta^{-}$-decay ($\beta$-decay) of 103 nuclei and $\beta^{+}$/EC decay of 99 nuclei, available for Z$>$82 in NUBASE2020 \cite{audi20201}. To estimate the precise half-lives of weak-decay, we have explored a few empirical formulas viz. Fiset and Nix formula \cite{Fiset1972}, Zhang formula \cite{zhang2006}, Modified Zhang formula \cite{zhang2007} and a recent formula modified by Sobhani \textit{et al} \cite{Sobhani} along with the results of half-lives obtained by using quasi-particle random-phase approximation (QRPA) \cite{frdm2019}. After this analysis for the nuclei Z $>$ 82 the results of Fiset and Nix formula are found with minimum RMSE which endorse the use of this formula for the estimation of half-lives of weak-decay in superheavy nuclei, as has been demonstrated in Refs. \cite{saxenaijmpe2019,singh2020,sharma2022}. The adopt formula is given by:
        \begin{equation}\label{tbeta}
         T_{\beta}(s)  = \frac{540 m_e^5}{\rho(W_\beta^6-m_e^6)}\times 10^{5.0}
        \end{equation}
This Eqn. (\ref{tbeta}) is only logical for $W_{\beta}\gg m_e$. For the average density of states $\rho$ in the daughter nucleus, we use the
empirical results given by Seeger \textit{et al} \cite{Seeger65}, from which $\rho$ has values 2.73 for even-even, 15 for odd-odd nuclei and 8.6 for odd $A$ (Mass number) nuclei, respectively. The formula for electron capture (EC) is given by:
\begin{eqnarray}
 T_{EC} (s) &=& \frac{9 m_e^2}{2\pi(\alpha Z_K)^{2s+1}\rho\left[Q_{EC}-(1-s)m_e\right]^3}\left(\frac{2R_0}{\hbar c/ m_e}\right)^{2-2s}
 \times\frac{\Gamma(2s+1)}{1+s}
 \nonumber\\ &&\times10^{6.5}
             \label{tecfinal}
\end{eqnarray}
In Eqn. (\ref{tecfinal}), $Z_K$ is the effective charge of the parent nucleus for an electron in the K-shell; it is approximately given by
$Z_K= Z_P - 0.35$. The energy $W_{\beta}$ is sum of energy of the emitted $\beta$-particle and its rest mass $m_e$, i.e. $W_{\beta} = Q_{\beta}+m_{e}$. Also, the quantity $s$ is given by $s = [ 1 -(\alpha Z_K)^2]^{\frac{1}{2}}$ and represents the rest mass of an electron minus its binding energy in the K-shell, in units of $m_e$. The quantity $\alpha$ is the fine-structure constant, and $R_0$ is the nuclear radius, which is taken to be $R_0 = 1.2249 A^{\frac{1}{3}} fm$.\par
\section{Result and discussion}
\subsection{$\alpha$-decay and decay chains}
\begin{figure}[!htbp]
\centering
\includegraphics[width=0.8\textwidth]{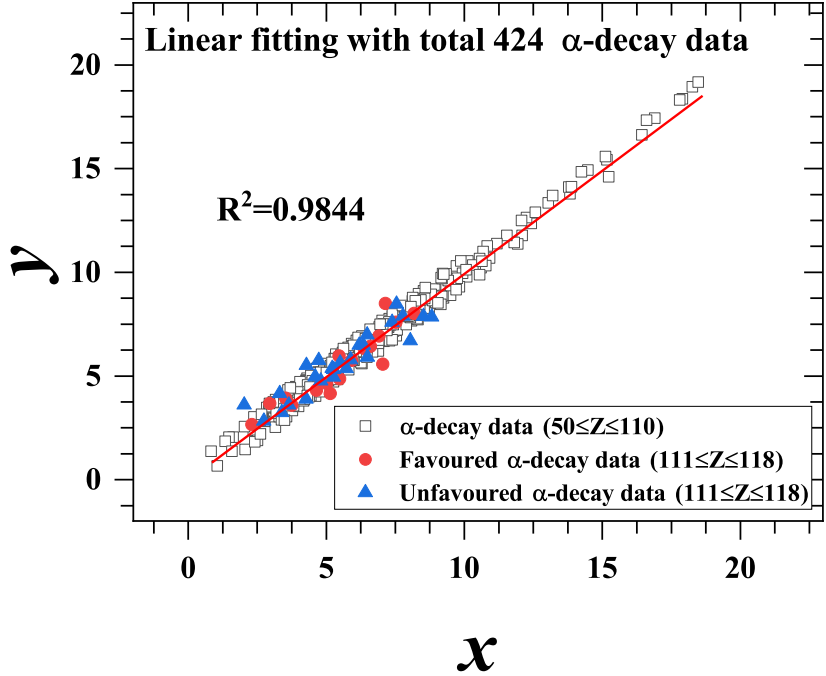}
\caption{(Colour online) Linear fitting using 424 experimental data in which x-axis
shows $x$=$(a\sqrt{\mu} + b)[(Z_{\alpha}Z_{d})^{0.6}Q_{\alpha}^{-1/2} - 7]$ and y-axis is $y$=$log_{10}T_{1/2}^{Exp.}-(c\sqrt{\mu} + d) - eI - fI^{2} - gl(l+1)$.}
\label{fig:ratio}
\end{figure}
In the present work, firstly we have looked into the $\alpha$-decay half-lives using the recent proposed NMHF formula \cite{sharma2021npa} mentioned in Eqn. (\ref{qf}) that has been found fairly accurate in the considered all the regions of periodic chart. We test this formula in considered superheavy region (with 17 favoured and 26 unfavoured available data for Z$>$110) in Fig. \ref{fig:ratio}, where we have plotted the variation in the quantity $y$=$log_{10}T_{1/2}^{Exp.}-(c\sqrt{\mu} + d) - eI - fI^{2} - gl(l+1)$ with the $Q_{\alpha}$ dependent terms of the formula for a total of 424 favoured and unfavoured $\alpha$-decay data. Almost all the points are found very near to the straight line which lead to R$^{2}$=0.9844 indicating a statistical measure of fit. It reflects an exemplary linear dependence of experimental $\alpha$-decay half-lives on $Q_{\alpha}$ dependent terms for the NMHF formula and it is quite satisfactory to note from Fig. \ref{fig:ratio} that the estimated half-lives for the considered range (for the isotopes from Rg to Og) match reasonably well with the experimental known half-lives. The points which are quite off in Fig. \ref{fig:ratio} are due to the large uncertainties in half-lives for example: $^{275,279}$Rg, $^{284,288}$Cn, $^{281,289}$Nh, $^{295}$Og etc. Hence, Fig. \ref{fig:ratio} vindicates the use of NMHF formula in this particular range of the nuclei (111$\leq$Z$\leq$118 and 161$\leq$N$\leq$192).

In this connection, we have estimated half-lives for the ground to ground favoured $\alpha$-transition (for which $l$ = 0) from Cn isotopes to Og isotopes which are shown in Table 2. $Q_{\alpha}$-values are taken from WS4 mass model \cite{ws42014} (mentioned in second and eighth column) due to its accuracy, judged over various other theories viz. relativistic mean-field theory (RMF) \cite{yadav2004,saxena2017,saxena2018,saxenaijmpe2019}, Finite Range Droplet Model (FRDM) \cite{moller2019data}, Relativistic continuum Hartree-Bogoliubov (RCHB) \cite{RCHB2018} etc. for almost 1500 nuclei in Ref. \cite{saxena2021}.
We have also calculated uncertainties in unknown $Q_{\alpha}$-values taken from WS4 mass model using Eqn. (\ref{uncer}) with the help of 69 experimental data in superheavy region \cite{audi20202}. Remarkably, the uncertainties in $Q_{\alpha}$-values are found only $\pm$0.04 MeV using which the half-lives are tabulated in sixth and twelfth column of the table along with their respective uncertainties.

 In a similar manner, the half-lives (including uncertainties) are estimated for ground to ground unfavoured $\alpha$-transition (for which $l$ $\neq$ 0) which are mentioned in Table \ref{tab:QF2}. The nuclei quoted in these tables (even-even and odd-A) are still undetected for which the theoretical spin parities for parent (j$_{p}$) and daughter (j$_{d}$) are taken from Ref. \cite{moller2019} From these Tables 2 and \ref{tab:QF2}, it is noticeable that estimated half-lives are within the experimental reach and expected to be useful for the future experiments.\par
\begin{table}[!htbp]
\begin{center}
\caption{$\alpha$-decay half-lives for ground to ground favoured transitions in undetected nuclei in the range 112$\leq$Z$\leq$118. These half-lives are calculated by NMHF formula for which the $Q_{\alpha}$-values and spin-parities are taken from Refs. \cite{ws42014} and \cite{moller2019}, respectively. The uncertainties in $Q_{\alpha}$-values are $\pm$0.04 MeV.}\end{center}
\def\arraystretch{1.15}
\resizebox{0.95\textwidth}{!}{%
{\begin{tabular}{c@{\hskip 0.2in}c@{\hskip 0.2in}c@{\hskip 0.2in}c@{\hskip 0.2in}c@{\hskip 0.2in}c|c@{\hskip
0.2in}c@{\hskip 0.2in}c@{\hskip 0.2in}c@{\hskip 0.3in}c@{\hskip 0.2in}c}
 \hline
\multicolumn{1}{c}{Favoured}&
\multicolumn{1}{c}{$Q_{\alpha}$}&
 \multicolumn{1}{c}{j$_{p}$}&
\multicolumn{1}{c}{j$_{d}$}&
\multicolumn{1}{c}{$l$}&
 \multicolumn{1}{c|}{$log_{10}T_{1/2}$}&
\multicolumn{1}{c}{Favoured}&
\multicolumn{1}{c}{$Q_{\alpha}$}&
 \multicolumn{1}{c}{j$_{p}$}&
\multicolumn{1}{c}{j$_{d}$}&
\multicolumn{1}{c}{$l$}&
 \multicolumn{1}{c}{$log_{10}T_{1/2}$}\\
\multicolumn{1}{c}{$\alpha$-Transition}&
\multicolumn{1}{c}{(MeV)}& \multicolumn{1}{c}{}& \multicolumn{1}{c}{}& \multicolumn{1}{c}{}& \multicolumn{1}{c|}{(s)}&
\multicolumn{1}{c}{$\alpha$-Transition}& \multicolumn{1}{c}{(MeV)}& \multicolumn{1}{c}{}& \multicolumn{1}{c}{}& \multicolumn{1}{c}{}&
\multicolumn{1}{c}{(s)}\\
 \hline
$^{274}$Cn$\rightarrow$$^{270}$Ds&11.55 &0$^{+}$   &0$^{+}$   &0&-3.83$\pm$0.08 & $^{296}$Cn$\rightarrow$$^{292}$Ds& 7.72 &0$^{+}$   &0$^{+}$  &0&6.53$\pm$0.14\\
$^{289}$Cn$\rightarrow$$^{285}$Ds& 9.05 &1/2$^{+}$ &1/2$^{+}$ &0& 2.22$\pm$0.11 & $^{297}$Cn$\rightarrow$$^{293}$Ds& 8.29 &1/2$^{+}$ &1/2$^{+}$&0&4.61$\pm$0.13\\
$^{290}$Cn$\rightarrow$$^{286}$Ds& 8.87 &0$^{+}$   &0$^{+}$   &0& 2.72$\pm$0.12 & $^{298}$Cn$\rightarrow$$^{294}$Ds& 8.79 &0$^{+}$   &0$^{+}$  &0&3.07$\pm$0.12\\
$^{291}$Cn$\rightarrow$$^{287}$Ds& 8.59 &1/2$^{+}$ &1/2$^{+}$ &0& 3.59$\pm$0.12 & $^{300}$Cn$\rightarrow$$^{296}$Ds& 8.43 &0$^{+}$   &0$^{+}$  &0&4.18$\pm$0.13\\
$^{292}$Cn$\rightarrow$$^{288}$Ds& 8.28 &0$^{+}$   &0$^{+}$   &0& 4.55$\pm$0.13 & $^{302}$Cn$\rightarrow$$^{298}$Ds& 7.51 &0$^{+}$   &0$^{+}$  &0&7.39$\pm$0.15\\
$^{294}$Cn$\rightarrow$$^{290}$Ds& 8.07 &0$^{+}$   &0$^{+}$   &0& 5.27$\pm$0.13 & $^{304}$Cn$\rightarrow$$^{300}$Ds& 7.31 &0$^{+}$   &0$^{+}$  &0&8.17$\pm$0.16\\
\hline
$^{275}$Nh$\rightarrow$$^{271}$Rg&11.93 &9/2$^{-}$ &9/2$^{-}$ &0&-4.36$\pm$0.07 & $^{277}$Nh$\rightarrow$$^{273}$Rg&12.20 &3/2$^{-}$&3/2$^{-}$&0&-4.80$\pm$0.07\\
\hline                                                                                                                                        $^{276}$Fl$\rightarrow$$^{272}$Cn&12.32 &0$^{+}$   &0$^{+}$   &0& -4.86$\pm$0.07& $^{296}$Fl$\rightarrow$$^{292}$Cn&8.56  &0$^{+}$   &0$^{+}$  &0&4.25$\pm$0.12\\
$^{278}$Fl$\rightarrow$$^{274}$Cn&12.52 &0$^{+}$   &0$^{+}$   &0& -5.17$\pm$0.07& $^{298}$Fl$\rightarrow$$^{294}$Cn&8.27  &0$^{+}$   &0$^{+}$  &0&5.21$\pm$0.13\\
$^{280}$Fl$\rightarrow$$^{276}$Cn&12.23 &0$^{+}$   &0$^{+}$   &0& -4.62$\pm$0.07& $^{300}$Fl$\rightarrow$$^{296}$Cn&9.56  &0$^{+}$   &0$^{+}$  &0&1.45$\pm$0.11\\
$^{282}$Fl$\rightarrow$$^{278}$Cn&11.38 &0$^{+}$   &0$^{+}$   &0& -2.98$\pm$0.08& $^{302}$Fl$\rightarrow$$^{298}$Cn&9.29  &0$^{+}$   &0$^{+}$  &0&2.20$\pm$0.11\\
$^{292}$Fl$\rightarrow$$^{288}$Cn& 8.95 &0$^{+}$   &0$^{+}$   &0&  3.01$\pm$0.12& $^{304}$Fl$\rightarrow$$^{300}$Cn&8.43  &0$^{+}$   &0$^{+}$  &0&4.77$\pm$0.13\\
$^{293}$Fl$\rightarrow$$^{289}$Cn& 8.78 &1/2$^{+}$ &1/2$^{+}$ &0&  3.52$\pm$0.12& $^{306}$Fl$\rightarrow$$^{302}$Cn&8.26  &0$^{+}$   &0$^{+}$  &0&5.35$\pm$0.13\\
$^{294}$Fl$\rightarrow$$^{290}$Cn& 8.71 &0$^{+}$   &0$^{+}$   &0&  3.75$\pm$0.12&                                  &  -   &    -     &   -   & -&  - \\
\hline
$^{297}$Mc$\rightarrow$$^{293}$Nh& 9.59 &5/2$^{-}$ &5/2$^{-}$ &0& 1.56$\pm$0.11 & $^{303}$Mc$\rightarrow$$^{299}$Nh&10.23 &5/2$^{-}$&5/2$^{-}$&0&0.02$\pm$0.10 \\
$^{299}$Mc$\rightarrow$$^{295}$Nh& 9.59 &5/2$^{-}$ &5/2$^{-}$ &0& 1.60$\pm$0.11 & $^{305}$Mc$\rightarrow$$^{301}$Nh& 9.31 &5/2$^{-}$&5/2$^{-}$&0&2.40$\pm$0.11 \\
$^{301}$Mc$\rightarrow$$^{297}$Nh&10.80 &5/2$^{-}$ &5/2$^{-}$ &0&-1.31$\pm$0.09 &                                  &  -   &      -   &     -   &-&  -   \\
\hline
$^{278}$Lv$\rightarrow$$^{274}$Fl&12.96&0$^{+}$   &0$^{+}$   &0& -5.58$\pm$0.07 & $^{296}$Lv$\rightarrow$$^{292}$Fl&10.89&0$^{+}$   &0$^{+}$  &0&-1.35$\pm$0.09 \\
$^{280}$Lv$\rightarrow$$^{276}$Fl&12.95&0$^{+}$   &0$^{+}$   &0& -5.52$\pm$0.07 & $^{298}$Lv$\rightarrow$$^{294}$Fl&10.77&0$^{+}$   &0$^{+}$  &0&-1.06$\pm$0.09 \\
$^{282}$Lv$\rightarrow$$^{278}$Fl&12.37&0$^{+}$   &0$^{+}$   &0& -4.50$\pm$0.07 & $^{300}$Lv$\rightarrow$$^{296}$Fl&10.92&0$^{+}$   &0$^{+}$  &0&-1.36$\pm$0.09 \\
$^{284}$Lv$\rightarrow$$^{280}$Fl&11.83&0$^{+}$   &0$^{+}$   &0& -3.46$\pm$0.08 & $^{301}$Lv$\rightarrow$$^{297}$Fl&11.58&1/2$^{+}$ &1/2$^{+}$&0&-2.73$\pm$0.08 \\
$^{286}$Lv$\rightarrow$$^{282}$Fl&11.31&0$^{+}$   &0$^{+}$   &0& -2.40$\pm$0.08 & $^{302}$Lv$\rightarrow$$^{298}$Fl&12.19&0$^{+}$   &0$^{+}$  &0&-3.89$\pm$0.07 \\
$^{287}$Lv$\rightarrow$$^{283}$Fl&11.28&3/2$^{+}$ &3/2$^{+}$ &0& -2.32$\pm$0.08 & $^{304}$Lv$\rightarrow$$^{300}$Fl&11.47&0$^{+}$   &0$^{+}$  &0&-2.47$\pm$0.08 \\
$^{288}$Lv$\rightarrow$$^{284}$Fl&11.29&0$^{+}$   &0$^{+}$   &0& -2.32$\pm$0.08 & $^{306}$Lv$\rightarrow$$^{302}$Fl&10.31&0$^{+}$   &0$^{+}$   &0&0.10$\pm$0.10 \\
$^{294}$Lv$\rightarrow$$^{290}$Fl&10.66&0$^{+}$   &0$^{+}$   &0& -0.88$\pm$0.09 & $^{308}$Lv$\rightarrow$$^{304}$Fl& 9.53&0$^{+}$  &0$^{+}$   &0& 2.08$\pm$0.11 \\
$^{295}$Lv$\rightarrow$$^{291}$Fl&10.77&1/2$^{+}$ &1/2$^{+}$ &0& -1.10$\pm$0.09 &                                  &   - &   -   &     -    &-&  -   \\
\hline
$^{280}$Og$\rightarrow$$^{276}$Lv&13.71 &0$^{+}$   &0$^{+}$   &0& -6.43$\pm$0.06& $^{296}$Og$\rightarrow$$^{292}$Lv&11.75 &0$^{+}$   &0$^{+}$  &0&-2.73$\pm$0.08\\
$^{282}$Og$\rightarrow$$^{278}$Lv&13.49 &0$^{+}$   &0$^{+}$   &0& -6.05$\pm$0.06& $^{297}$Og$\rightarrow$$^{293}$Lv&12.10 &1/2$^{+}$ &1/2$^{+}$&0&-3.40$\pm$0.08\\
$^{284}$Og$\rightarrow$$^{280}$Lv&13.23 &0$^{+}$   &0$^{+}$   &0& -5.58$\pm$0.07& $^{298}$Og$\rightarrow$$^{294}$Lv&12.18 &0$^{+}$   &0$^{+}$  &0&-3.53$\pm$0.07\\
$^{285}$Og$\rightarrow$$^{281}$Lv&13.07 &1/2$^{+}$ &1/2$^{+}$ &0& -5.31$\pm$0.07& $^{300}$Og$\rightarrow$$^{296}$Lv&11.96 &0$^{+}$   &0$^{+}$  &0&-3.08$\pm$0.08\\
$^{286}$Og$\rightarrow$$^{282}$Lv&12.92 &0$^{+}$   &0$^{+}$   &0& -5.03$\pm$0.07& $^{302}$Og$\rightarrow$$^{298}$Lv&12.04 &0$^{+}$   &0$^{+}$  &0&-3.21$\pm$0.08\\
$^{287}$Og$\rightarrow$$^{283}$Lv&12.80 &1/2$^{+}$ &1/2$^{+}$ &0& -4.81$\pm$0.07& $^{304}$Og$\rightarrow$$^{300}$Lv&13.12 &0$^{+}$   &0$^{+}$  &0&-5.13$\pm$0.07\\
$^{288}$Og$\rightarrow$$^{284}$Lv&12.62 &0$^{+}$   &0$^{+}$   &0& -4.47$\pm$0.07& $^{306}$Og$\rightarrow$$^{302}$Lv&12.48 &0$^{+}$   &0$^{+}$  &0&-3.99$\pm$0.07\\
$^{289}$Og$\rightarrow$$^{285}$Lv&12.59 &3/2$^{+}$ &3/2$^{+}$ &0& -4.41$\pm$0.07& $^{308}$Og$\rightarrow$$^{304}$Lv&11.20 &0$^{+}$   &0$^{+}$  &0&-1.47$\pm$0.09\\
$^{290}$Og$\rightarrow$$^{286}$Lv&12.60 &0$^{+}$   &0$^{+}$   &0& -4.41$\pm$0.07& $^{310}$Og$\rightarrow$$^{306}$Lv&10.43 &0$^{+}$   &0$^{+}$  &0&0.29$ \pm$0.10\\
$^{292}$Og$\rightarrow$$^{288}$Lv&12.24 &0$^{+}$   &0$^{+}$   &0& -3.73$\pm$0.07&                                 &  -   &    -     &    -     &-& -   \\
 \hline
\end{tabular}}
}
\label{NMHF}

\end{table}
\begin{table}[!htbp]
\caption{Same as Table 2 but for unfavoured transitions in the range 111$\leq$Z$\leq$118.}
\centering
\def\arraystretch{1.15}
\resizebox{0.95\textwidth}{!}{%
{\begin{tabular}{c@{\hskip 0.2in}c@{\hskip 0.2in}c@{\hskip 0.2in}c@{\hskip 0.2in}c@{\hskip 0.2in}c|c@{\hskip
0.2in}c@{\hskip 0.2in}c@{\hskip 0.2in}c@{\hskip 0.3in}c@{\hskip 0.2in}c}
 \hline
\multicolumn{1}{c}{Unfavoured}&
\multicolumn{1}{c}{$Q_{\alpha}$}&
 \multicolumn{1}{c}{j$_{p}$}&
\multicolumn{1}{c}{j$_{d}$}&
\multicolumn{1}{c}{$l$}&
 \multicolumn{1}{c|}{$log_{10}T_{1/2}$}&
\multicolumn{1}{c}{Unfavoured}&
\multicolumn{1}{c}{$Q_{\alpha}$}&
 \multicolumn{1}{c}{j$_{p}$}&
\multicolumn{1}{c}{j$_{d}$}&
\multicolumn{1}{c}{$l$}&
 \multicolumn{1}{c}{$log_{10}T_{1/2}$}\\
\multicolumn{1}{c}{$\alpha$-Transition}&
\multicolumn{1}{c}{(MeV)}& \multicolumn{1}{c}{}& \multicolumn{1}{c}{}& \multicolumn{1}{c}{}& \multicolumn{1}{c|}{(s)}&
\multicolumn{1}{c}{$\alpha$-Transition}& \multicolumn{1}{c}{(MeV)}& \multicolumn{1}{c}{}& \multicolumn{1}{c}{}& \multicolumn{1}{c}{}&
\multicolumn{1}{c}{(s)}\\
 \hline
$^{287}$Rg$\rightarrow$$^{283}$Mt&8.44 &13/2$^{+}$  &9/2$^{-}$ &3&4.02$\pm$0.12& $^{297}$Rg$\rightarrow$$^{293}$Mt&8.38 &3/2$^{-}$  &1/2$^{-}$&2&4.20$\pm$0.13 \\
$^{289}$Rg$\rightarrow$$^{285}$Mt&8.30 &3/2$^{-}$  &9/2$^{-}$  &4&4.68$\pm$0.13& $^{299}$Rg$\rightarrow$$^{295}$Mt&7.85 &3/2$^{+}$  &5/2$^{+}$&2&5.96$\pm$0.14 \\
$^{291}$Rg$\rightarrow$$^{287}$Mt&7.87 &3/2$^{-}$  &1/2$^{+}$  &1&5.71$\pm$0.14& $^{301}$Rg$\rightarrow$$^{297}$Mt&7.15 &3/2$^{+}$  &5/2$^{+}$&2&8.61$\pm$0.16 \\
$^{293}$Rg$\rightarrow$$^{289}$Mt&7.75 &3/2$^{-}$  &5/2$^{-}$  &2&6.23$\pm$0.14& $^{303}$Rg$\rightarrow$$^{299}$Mt&6.85 &3/2$^{+}$  &5/2$^{+}$&2&9.87$\pm$0.17 \\
$^{295}$Rg$\rightarrow$$^{291}$Mt&7.40 &3/2$^{-}$  &1/2$^{-}$  &2&7.56$\pm$0.15&                                  &  -  &    -   &     -     &-&  -  \\
\hline
$^{273}$Cn$\rightarrow$$^{269}$Ds&11.64&3/2$^{+}$  &1/2$^{+}$  &2&-3.88$\pm$0.08& $^{299}$Cn$\rightarrow$$^{295}$Ds&8.69 &3/2$^{+}$&1/2$^{+}$&2&3.44$^{+0.02}_{-0.22}$ \\
$^{275}$Cn$\rightarrow$$^{271}$Ds&11.74&13/2$^{-}$ &3/2$^{+}$  &5&-3.46$\pm$0.08& $^{301}$Cn$\rightarrow$$^{297}$Ds&7.81 &5/2$^{+}$&3/2$^{+}$&2&6.31$^{+0.05}_{-0.24}$ \\
$^{293}$Cn$\rightarrow$$^{289}$Ds&8.19 &3/2$^{+}$  &1/2$^{+}$  &2& 5.02$\pm$0.13& $^{303}$Cn$\rightarrow$$^{299}$Ds&7.44 &7/2$^{+}$&5/2$^{+}$&2&7.71$^{+0.06}_{-0.25}$ \\
$^{295}$Cn$\rightarrow$$^{291}$Ds&7.86 &1/2$^{+}$  &3/2$^{+}$  &2& 6.14$\pm$0.14&                                  &  -  &   -    &    -    &-& -   \\
\hline
$^{291}$Nh$\rightarrow$$^{287}$Rg& 8.91 &1/2$^{-}$ &13/2$^{+}$&7&4.22$\pm$0.12& $^{299}$Nh$\rightarrow$$^{295}$Rg& 9.13 &5/2$^{-}$&3/2$^{-}$&2&2.49$\pm$0.11  \\
$^{293}$Nh$\rightarrow$$^{289}$Rg& 8.48 &7/2$^{-}$ &3/2$^{-}$ &2&4.34$\pm$0.13& $^{301}$Nh$\rightarrow$$^{297}$Rg& 8.83 &7/2$^{-}$&3/2$^{-}$&2&3.38$\pm$0.12  \\
$^{295}$Nh$\rightarrow$$^{291}$Rg& 8.15 &7/2$^{-}$ &3/2$^{-}$ &2&5.42$\pm$0.13& $^{303}$Nh$\rightarrow$$^{299}$Rg& 7.87 &1/2$^{+}$&3/2$^{+}$&2&6.52$\pm$0.14  \\
$^{297}$Nh$\rightarrow$$^{293}$Rg& 7.88 &5/2$^{-}$ &3/2$^{-}$ &2&6.39$\pm$0.14& $^{305}$Nh$\rightarrow$$^{301}$Rg& 7.80 &1/2$^{+}$&3/2$^{+}$&2&6.79$\pm$0.14  \\
\hline
$^{275}$Fl$\rightarrow$$^{271}$Cn&12.51 &3/2$^{+}$ &1/2$^{+}$ &2&-5.06$\pm$0.07& $^{297}$Fl$\rightarrow$$^{293}$Cn&8.35  &1/2$^{+}$ &3/2$^{+}$&2&5.05$\pm$0.13 \\
$^{277}$Fl$\rightarrow$$^{273}$Cn&12.40 &13/2$^{-}$&3/2$^{+}$ &5&-4.27$\pm$0.07& $^{299}$Fl$\rightarrow$$^{295}$Cn&8.91  &11/2$^{-}$&1/2$^{+}$&5&3.36$^{+0.48}_{-0.71}$ \\
$^{279}$Fl$\rightarrow$$^{275}$Cn&12.43 &5/2$^{+}$ &13/2$^{-}$&5&-4.28$\pm$0.07& $^{301}$Fl$\rightarrow$$^{297}$Cn&9.58  &1/2$^{-}$ &1/2$^{+}$&1&1.43$^{+0.08}_{-0.13}$ \\
$^{281}$Fl$\rightarrow$$^{277}$Cn&11.82 &1/2$^{+}$ &3/2$^{+}$ &2&-3.70$\pm$0.08& $^{303}$Fl$\rightarrow$$^{299}$Cn&8.95  &5/2$^{+}$ &3/2$^{+}$&2&3.21$^{+0.02}_{-0.21}$ \\
$^{283}$Fl$\rightarrow$$^{279}$Cn&10.88 &3/2$^{+}$ &11/2$^{+}$&4&-1.45$\pm$0.09& $^{305}$Fl$\rightarrow$$^{301}$Cn&8.28  &9/2$^{-}$ &5/2$^{+}$&3&5.32$^{+0.08}_{-0.35}$ \\
$^{295}$Fl$\rightarrow$$^{291}$Cn&8.61  &3/2$^{+}$ &1/2$^{+}$ &2& 4.23$\pm$0.12&                                  &   -  &    -     &    -     &-& -  \\
\hline
$^{277}$Mc$\rightarrow$$^{273}$Nh&12.64 &1/2$^{-}$ &3/2$^{-}$ &2& -5.09$\pm$0.07& $^{285}$Mc$\rightarrow$$^{281}$Nh&10.73 &1/2$^{-}$ &7/2$^{-}$&4&-0.90$\pm$0.09\\
$^{279}$Mc$\rightarrow$$^{275}$Nh&12.65 &1/2$^{-}$ &9/2$^{-}$ &4& -4.73$\pm$0.07& $^{293}$Mc$\rightarrow$$^{289}$Nh&9.71  &5/2$^{-}$ &7/2$^{-}$&2&1.32 $\pm$0.10\\
$^{281}$Mc$\rightarrow$$^{277}$Nh&12.20 &1/2$^{-}$ &3/2$^{-}$ &2& -4.24$\pm$0.07& $^{295}$Mc$\rightarrow$$^{291}$Nh&9.72  &5/2$^{-}$ &1/2$^{-}$&2&1.33 $\pm$0.10\\
$^{283}$Mc$\rightarrow$$^{279}$Nh&11.32 &1/2$^{-}$ &3/2$^{-}$ &2& -2.52$\pm$0.08& $^{307}$Mc$\rightarrow$$^{303}$Nh&8.90  &5/2$^{-}$ &1/2$^{+}$&3& 3.89$\pm$0.12\\
\hline
$^{277}$Lv$\rightarrow$$^{273}$Fl&13.12 &3/2$^{+}$ &1/2$^{+}$ &2&-5.72$\pm$0.07& $^{297}$Lv$\rightarrow$$^{293}$Fl&10.84 &3/2$^{+}$&1/2$^{+}$&2&-1.07$\pm$0.09  \\
$^{279}$Lv$\rightarrow$$^{275}$Fl&13.02 &13/2$^{-}$&3/2$^{+}$ &5&-4.95$\pm$0.07& $^{299}$Lv$\rightarrow$$^{295}$Fl&10.84 &1/2$^{+}$&3/2$^{+}$&2&-1.05$\pm$0.09  \\
$^{281}$Lv$\rightarrow$$^{277}$Fl&12.70 &1/2$^{+}$ &13/2$^{-}$&7&-3.76$\pm$0.07& $^{303}$Lv$\rightarrow$$^{299}$Fl&11.93&3/2$^{+}$&11/2$^{-}$&5&-3.27$^{+0.52}_{-0.67}$  \\
$^{283}$Lv$\rightarrow$$^{279}$Fl&12.11 &1/2$^{+}$ &5/2$^{+}$ &2&-3.85$\pm$0.07& $^{305}$Lv$\rightarrow$$^{301}$Fl&10.88 &5/2$^{+}$&1/2$^{-}$&3&-1.15$^{+0.13}_{-0.30}$  \\
$^{285}$Lv$\rightarrow$$^{281}$Fl&11.55 &3/2$^{+}$ &1/2$^{+}$ &2&-2.75$\pm$0.08& $^{307}$Lv$\rightarrow$$^{303}$Fl&9.84  &7/2$^{+}$ &5/2$^{+}$&2&1.30$^{+0.01}_{-0.20}$  \\
\hline
$^{279}$Ts$\rightarrow$$^{275}$Mc&13.34&3/2$^{-}$ &1/2$^{-}$ &2&-5.88$\pm$0.06& $^{297}$Ts$\rightarrow$$^{293}$Mc&11.62&3/2$^{-}$&5/2$^{-}$&2&-2.51$\pm$0.08  \\
$^{281}$Ts$\rightarrow$$^{277}$Mc&13.21&3/2$^{-}$ &1/2$^{-}$ &2&-5.64$\pm$0.06& $^{299}$Ts$\rightarrow$$^{295}$Mc&11.46&1/2$^{-}$&5/2$^{-}$&2&-2.16$\pm$0.08  \\
$^{283}$Ts$\rightarrow$$^{279}$Mc&12.88&3/2$^{-}$ &1/2$^{-}$ &2&-5.05$\pm$0.07& $^{301}$Ts$\rightarrow$$^{297}$Mc&11.61&1/2$^{-}$&5/2$^{-}$&2&-2.44$\pm$0.08 \\
$^{285}$Ts$\rightarrow$$^{281}$Mc&12.44&3/2$^{-}$ &1/2$^{-}$ &2&-4.25$\pm$0.07& $^{303}$Ts$\rightarrow$$^{299}$Mc&12.78&1/2$^{-}$&5/2$^{-}$&2&-4.59$\pm$0.07 \\
$^{287}$Ts$\rightarrow$$^{283}$Mc&12.05&3/2$^{-}$ &1/2$^{-}$ &2&-3.50$\pm$0.07& $^{305}$Ts$\rightarrow$$^{301}$Mc&12.08&1/2$^{-}$&5/2$^{-}$&2&-3.31$\pm$0.08 \\
$^{289}$Ts$\rightarrow$$^{285}$Mc&11.99&3/2$^{-}$ &1/2$^{-}$ &2&-3.34$\pm$0.08& $^{307}$Ts$\rightarrow$$^{303}$Mc&10.92&5/2$^{+}$&5/2$^{-}$&1&-1.02$\pm$0.09  \\
$^{295}$Ts$\rightarrow$$^{291}$Mc&11.30&3/2$^{-}$ &5/2$^{-}$ &2&-1.88$\pm$0.08& $^{309}$Ts$\rightarrow$$^{305}$Mc&10.01&9/2$^{+}$ &5/2$^{-}$&3&1.38$\pm$0.10 \\
\hline
$^{279}$Og$\rightarrow$$^{275}$Lv&13.78 &3/2$^{+}$  &1/2$^{+}$ &2&-6.42$\pm$0.06 &$^{301}$Og$\rightarrow$$^{297}$Lv&12.02 &1/2$^{+}$ &3/2$^{+}$&2&-3.04$\pm$0.08\\
$^{281}$Og$\rightarrow$$^{277}$Lv&13.76 &13/2$^{-}$ &3/2$^{+}$ &5&-5.78$\pm$0.06 &$^{303}$Og$\rightarrow$$^{299}$Lv&12.61 &7/2$^{+}$ &1/2$^{+}$&4&-4.15$^{+0.31}_{-0.45}$ \\
$^{283}$Og$\rightarrow$$^{279}$Lv&13.33 &1/2$^{+}$  &13/2$^{-}$&7&-4.44$\pm$0.06 &$^{305}$Og$\rightarrow$$^{301}$Lv&12.91 &5/2$^{+}$ &1/2$^{+}$&2&-4.71$^{+0.03}_{-0.16}$ \\
$^{291}$Og$\rightarrow$$^{287}$Lv&12.42 &5/2$^{+}$  &3/2$^{+}$ &2&-3.93$\pm$0.07 &$^{307}$Og$\rightarrow$$^{303}$Lv&11.92 &5/2$^{+}$ &3/2$^{+}$&2&-2.89$^{+0.02}_{-0.17}$ \\
$^{293}$Og$\rightarrow$$^{289}$Lv&12.24 &1/2$^{+}$  &5/2$^{+}$ &2&-3.57$\pm$0.07 &$^{309}$Og$\rightarrow$$^{305}$Lv&10.72 &7/2$^{+}$ &5/2$^{+}$&2&-0.34$^{+0.01}_{-0.19}$ \\
$^{299}$Og$\rightarrow$$^{295}$Lv&12.05 &3/2$^{+}$  &1/2$^{+}$ &2&-3.12$\pm$0.08 &                                 &   -  &      -    &    -   &-&   - \\
\hline
\end{tabular}}
}
\label{tab:QF2}
\end{table}
\begin{figure*}[!htbp]
\centering
\includegraphics[width=1.00\textwidth]{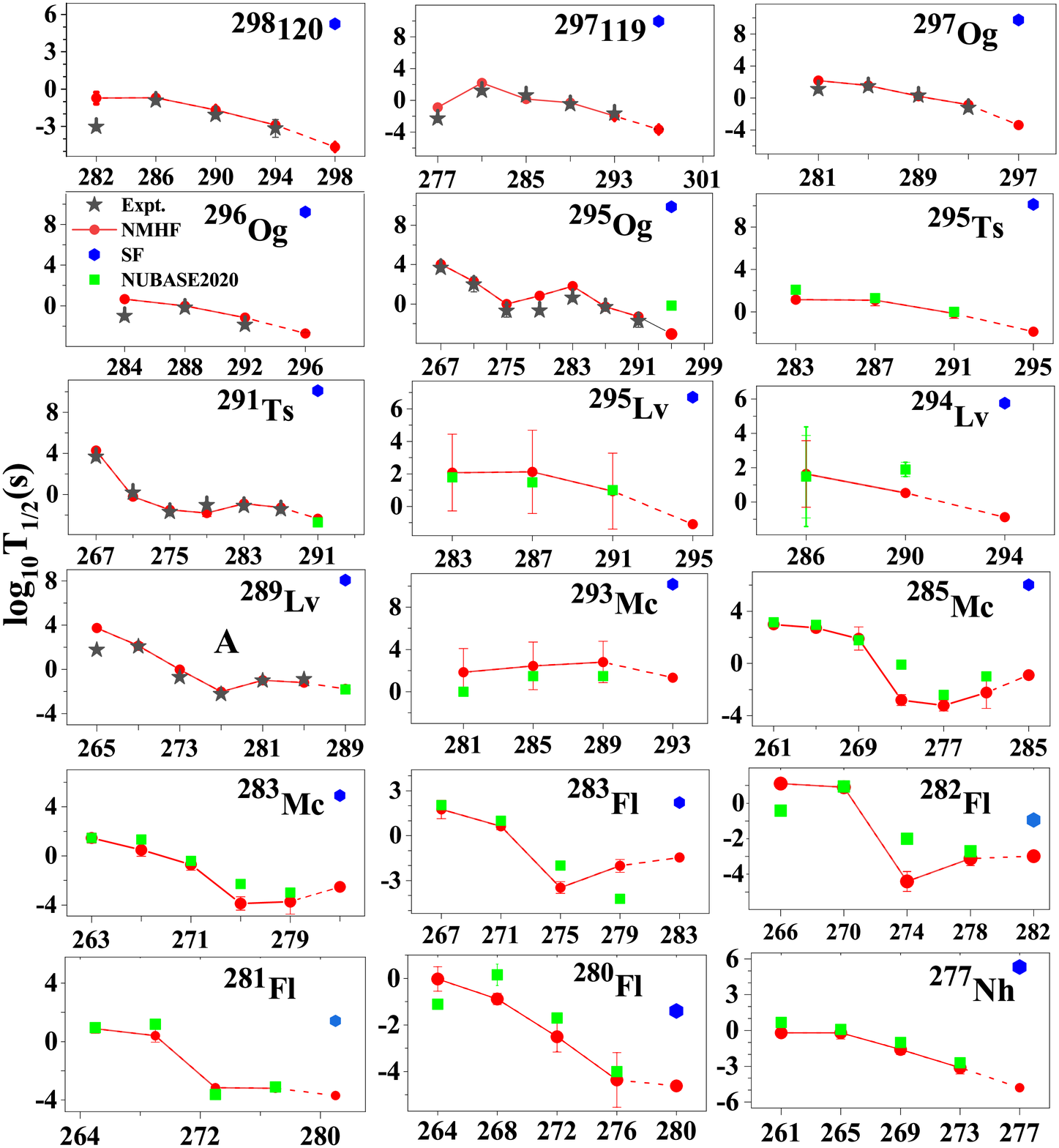}
\caption{(Colour online) $\alpha$-decay half-lives for experimental decay chains calculated by the NMHF formula (red circles) are shown. Experimental
data (black stars) are taken from Refs. \cite{og2017,og1999,og2015npa,forsberg2016,og2000} whereas the latest evaluated data (green squares) are taken from NUBASE2020 \cite{audi20201}. Error bars are also mentioned in the graph: for both theoretical and experimental values. For the probable candidate (connected by dash lines), SF half-lives are depicted by blue hexagons.}
\label{fig:decay-chain-exp}
\end{figure*}
\begin{figure*}[!htbp]
\centering
\includegraphics[width=0.90\textwidth]{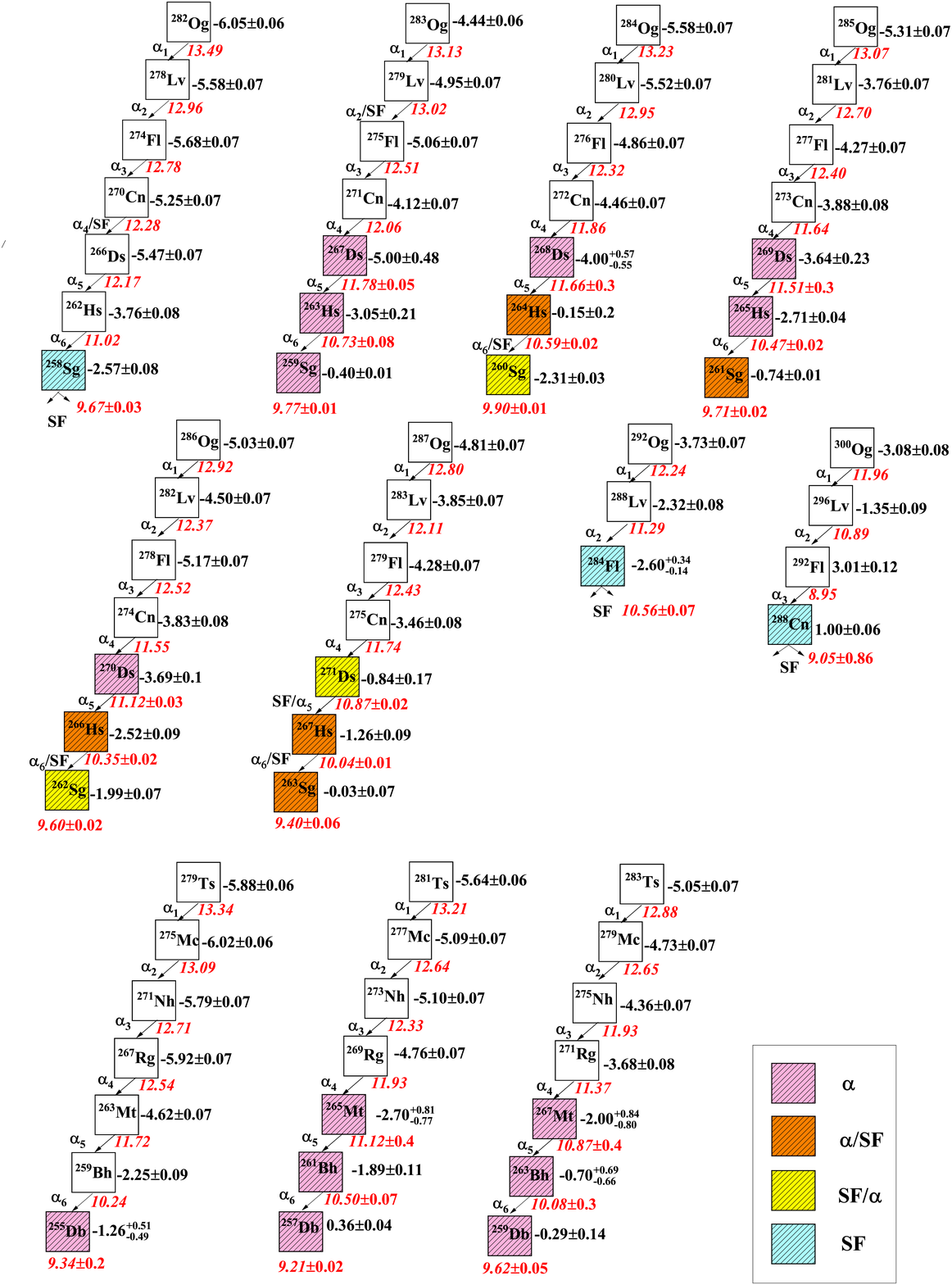}
\caption{(Colour online) Theoretical decay chains by using NMHF formula (Black numbers indicate the logarithmic half-lives with uncertainties in second). Red numbers represent the $Q_{\alpha}$-values (in MeV) taken from WS4 mass model \cite{ws42014} or AME2020 \cite{audi20202}. The uncertainties in theoretical $Q_{\alpha}$-values is $\pm$0.04 MeV and the shaded blocks show the modes which are available in NUBASE2020 \cite{audi20201}.}
\label{fig:thchains}
\end{figure*}

As mentioned above, detection of SHN is mainly governed by observation of $\alpha$-decay chains, therefore, matching of theoretical calculations with already observed decay chains is a prerequisite before estimating theoretical $\alpha$-decay chains for undetected nuclei. With this in view, in Fig. \ref{fig:decay-chain-exp}, we have compared $\alpha$-decay half-lives from the known decay chains \cite{og2017,og1999,og2015npa,forsberg2016,og2000} of various nuclei in between Rg and Og elements. The theoretical half-lives from the NMHF formula are found in excellent match which once again demonstrate its accuracy. The error bars are also shown with the theoretical half-lives by using the uncertainties in experimental $Q_{\alpha}$-values \cite{og2017,og1999,og2015npa,forsberg2016,og2000}. Additionally, in the each panel of Fig. \ref{fig:decay-chain-exp} half-life of most probable nucleus adjacent to the known decay chains \cite{og2017,og1999,og2015npa,forsberg2016,og2000} (shown by star symbols) or known nuclei as per NUBASE2020 \cite{audi20201} (shown by square symbols) is shown by connecting dotted lines. The SF half-life calculated by using Eqn. (\ref{baoSF}) for next probable nucleus is also depicted by blue circle which indicates quite larger value comparative to $\alpha$-decay half-life, and hence signifies the greater probability of $\alpha$-decay as comparison to SF in such nucleus. These predictions leading to the potential upper parts of known decay chains are also found consistent with the estimation by NUBASE2020 \cite{audi20201}. Therefore, most probable nuclei which are proposed here for the future detection in relation with known decay chains are: $^{298}$120, $^{297}$119, $^{297,296,295}$Og, $^{295,291}$Ts, $^{295,294,289}$Lv, $^{293,285,283}$Mc, $^{283,282,281,280}$Fl and $^{277}$Nh. Possible reactions resulting these probable candidates along with cross-section will be discussed in our subsequent article.\par

There is more possible expansion of regions belonging to $\alpha$-decay as well as potential $\alpha$-decay chains which can also be investigated
within the same approach as mentioned above. With this in view, we have calculated $\alpha$-decay half-lives (using Eqn. (\ref{qf})) in conjunction with SF half-lives (using Eqn. (\ref{baoSF})) for the nuclei within 111$\leq$Z$\leq$118 and 161$\leq$N$\leq$192. Both half-lives are calculated for various possible decay chains for the mentioned range and are tabulated in Appendix for the readers. The uncertainties in theoretical half-lives of $\alpha$-decay are calculated with the help of uncertainties in theoretical $Q_{\alpha}$-values of WS4 ($\pm$0.04 MeV). Many of the theoretical $\alpha$-decay chains are found with 3 $\alpha$-transitions or more. Gratifyingly, $\alpha$-decay chain of $^{281}$Og is found with 6 $\alpha$-transitions which leads to the adequate possibility of detection of these sets of nuclei in future experiments. Few selective decay chains in which one or more terminating nuclei are known as per NUBASE2020 \cite{audi20201}, are shown in Fig. \ref{fig:thchains}. These decay chains mainly consist of nuclei between Og (Z$=$118) to Db (Z$=$105). So, theoretical estimates lead to $\alpha$-decay chains of $^{282-287,292,300}$Og (rest others in between 288$\leq$A$\leq$291 and 293$\leq$A$\leq$299 are already part of Fig. \ref{fig:decay-chain-exp}) which are represented graphically in Fig. \ref{fig:thchains}. In a similar way, the $\alpha$-decay chains of $^{279-283}$Ts (odd-A) are shown. In Fig. \ref{fig:thchains}, $Q_\alpha$-values (red numbers) are taken from AME2020 \cite{audi20202} wherever available, otherwise from WS4 mass model \cite{ws42014}. The black numbers indicate logarithmic half-lives estimated by NMHF formula. The shaded blocks show the modes which are available in NUBASE2020 \cite{audi20201} and the colour of blocks corresponds to $\alpha$, SF, $\alpha$/SF or SF/$\alpha$. Most of the decay chains presented in Fig. \ref{fig:thchains} comprise several already known nuclei and, therefore, the nuclei with theoretical estimates in these decay chains are likely to be observed experimentally in future.\par

\subsection{Cluster Decay}
\begin{figure*}[!htbp]
\centering
\includegraphics[width=0.85\textwidth]{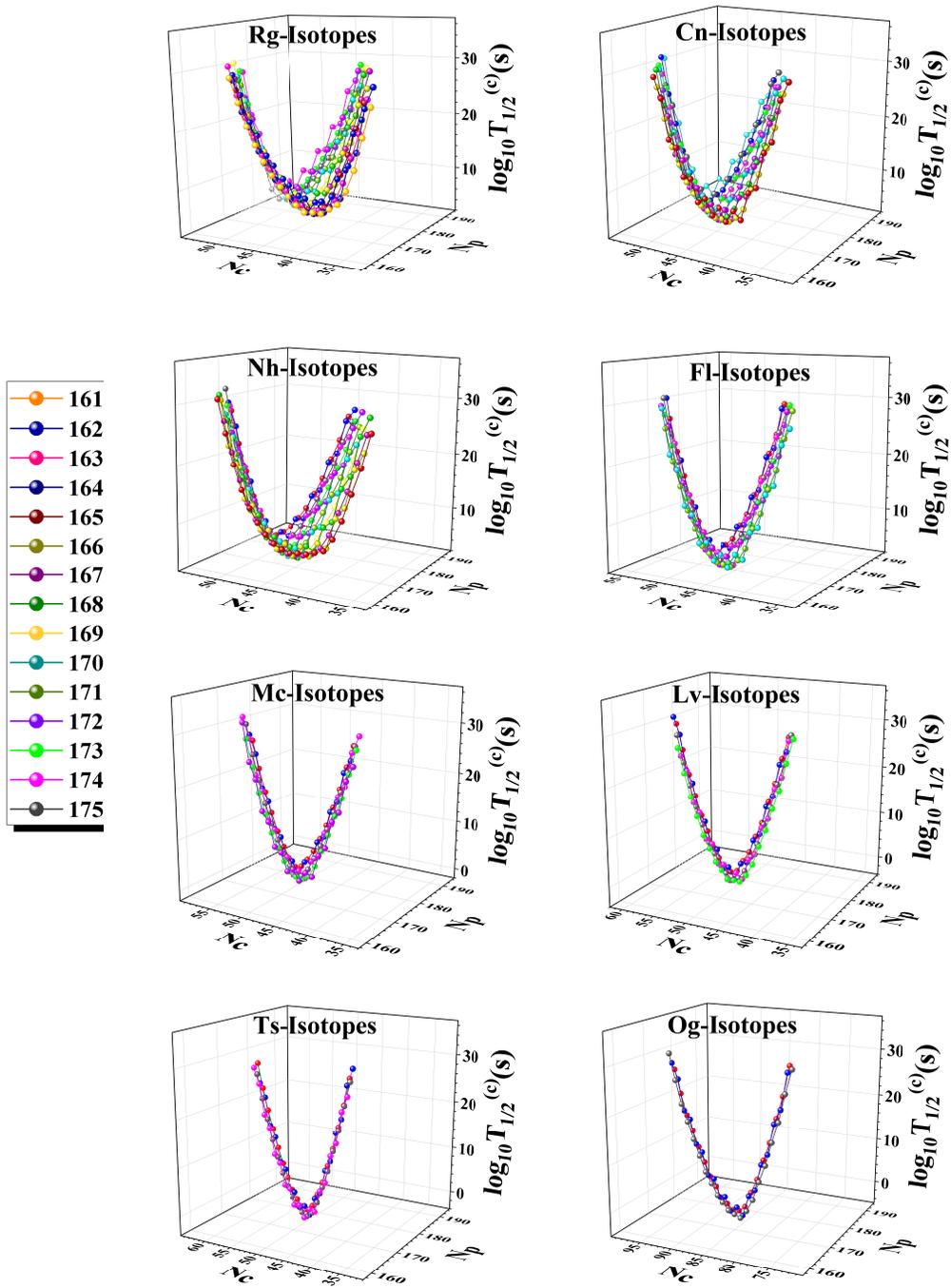}
\caption{(Colour online) Half-lives of cluster emission (in second) from experimentally known Rg to Og isotopes (with N$=$161-177) by using UDL formula \cite{udl2009}. N$_p$ and N$_c$ represent neutron numbers of parent nuclei (Rg, Cn, Nh, Fl, Mc, Lv, Ts and Og) and respective clusters (Cu, Zn, Ga, Ge, As, Se, Br and Kr).}
\label{fig:clusters}
\end{figure*}
In the following, we have probed cluster decay from the mentioned SHN by calculating the logarithmic half-lives using UDL formula \cite{udl2009}.
This formula is given in Eqn. (\ref{udl}) which mainly relies on $Q$-value calculated by using the following relation:
\begin{eqnarray}
 Q (MeV) &=& B.E.(d) + B.E.(c) - B.E.(p)+  k[Z_{p}^{\beta}- Z_{d}^{\beta}]
 \label{Q}
 \end{eqnarray}
The term $k[Z_{p}^{\epsilon}-Z_{d}^{\epsilon}]$ indicates the screening effect caused by the surrounding electrons around the nucleus \cite{Denisov2009prc} where k=8.7 eV [8.7 $\times$ 10$^{-6}$MeV] and $\epsilon$=2.517 for Z (proton number) $\geq$ 60 and k=13.6 eV [13.6 $\times$ 10$^{-6}$MeV] and $\epsilon$ =2.408 for Z $<$ 60 have been deducted from the data shown by Huang \textit{et al} \cite{huang1976}. For this study, we have taken experimenral binding energies (for daughter(d), cluster(c), and parent(p) nuclei) from AME2020 \cite{audi20202} to calculate $Q$-values for cluster emission, considering the fact that the validity of this mass model for the cluster emission in this superheavy region has already been verified in Refs. \cite{santhosh2018,soylu2019}. We have also calculated the uncertainties in these $Q$-values: $\pm$0.05 MeV, with the help of Eqn. (\ref{uncer}) and using 73 experimental data of $Q$-values for cluster emission \cite{barwick1986,bonetti2007}.\par

In Fig. \ref{fig:clusters}, we have shown half-lives of cluster emission from experimentally known Rg to Og isotopes (with N$=$161-177) in various panels, respectively. Considering Z$_{d}$ = 82 shell closure effect \cite{Andreyev2013}, i.e. the daughter as one of the isotopes of Pb, the isotopes of Cu, Zn, Ga, Ge, As, Se, Br, and Kr are taken into account as probable clusters from the parent isotopes of Rg, Cn, Nh, Fl, Mc, Lv, Ts and Og, respectively. From Fig. \ref{fig:clusters}, parabolic trend is clearly seen where each parabola represents half-lives of one cluster from the corresponding chain of parent isotopes. Half-lives are shown up to $log_{10}T_c$=30 s considering the experimental limit of cluster emission \cite{bonetti2007}. From this extensive analysis, one can find a minima of each parabola which represents the potential cluster nucleus for a particular parent nucleus. From Fig. \ref{fig:clusters}, one can notice that the half-life of minima of each parabola lies somewhat close to the range of half-life of $\alpha$-decay as can be seen from Figs. \ref{fig:decay-chain-exp}, \ref{fig:thchains} and Tables 2, \ref{tab:QF2}.  Therefore, cluster decay is also found probable provided that it invariably competes with $\alpha$ and SF decays in the superheavy region.

\subsection{Competition among various decay-modes}
In the preceding sections, $\alpha$-decay, spontaneous fission as well as cluster decay have been discussed for the considered range of nuclei.
Competition among these decay modes is crucial to investigate the possibility of synthesis of new isotopes as well as to visualize prospects of other decay modes in known nuclei. Before probing the competition among mentioned decay modes, it will be worth to include weak-decay mode in this contest as it has already been conjectured as one of the possible decay modes in the superheavy region, though in an exceptional manner \cite{singh2020,sarriguren2019,sarriguren2018}.\par
\begin{figure*}[!htbp]
\centering
\includegraphics[width=0.85\textwidth]{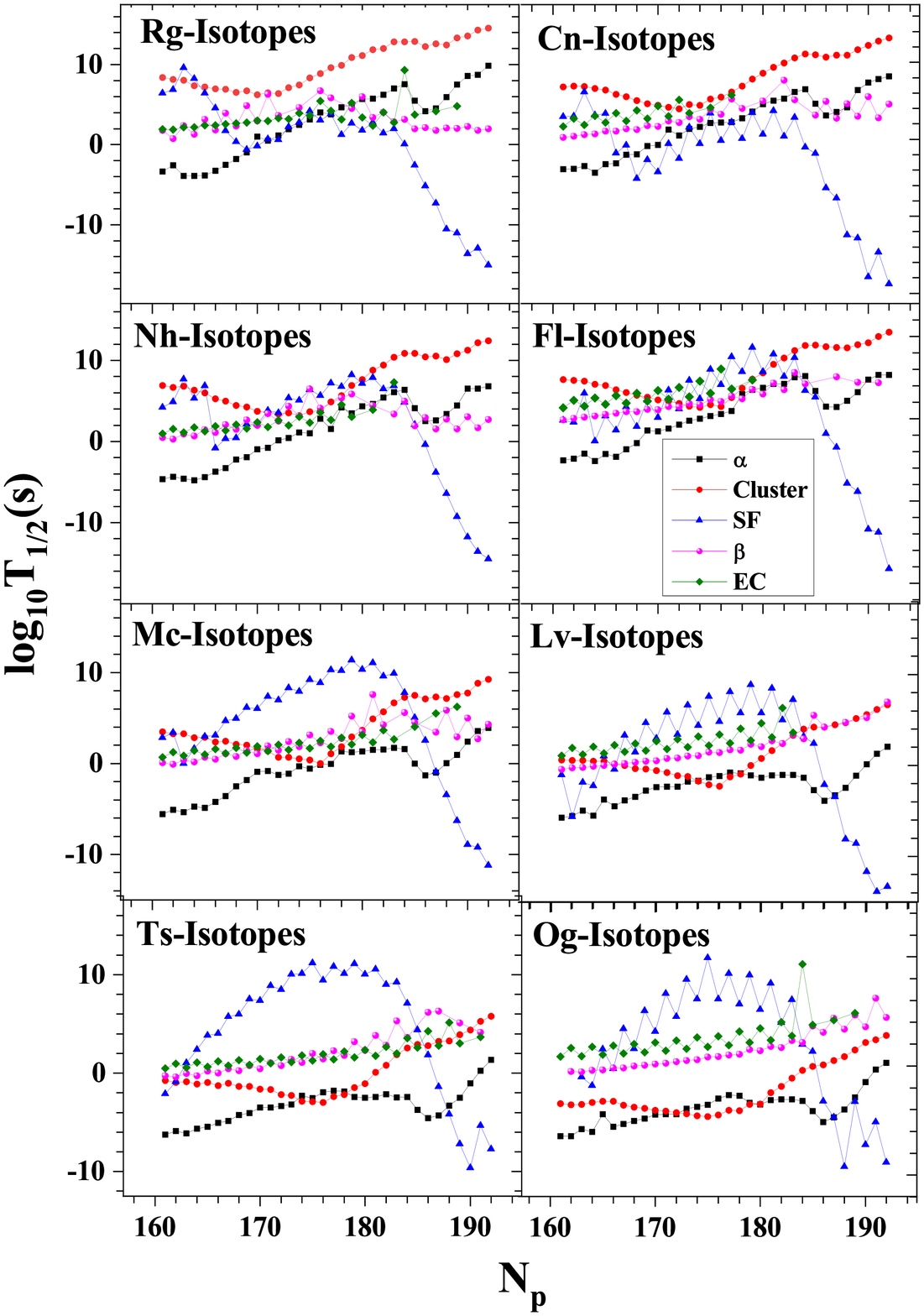}
\caption{(Colour online) Competition among the various decay modes in the range 111$\leq$Z$\leq$118. Half-lives for these decay modes are shown as a function of the neutron number for isotopes from Rg to Og.}
\label{fig:decaymodes}
\end{figure*}
{\small
\begin{center}
\begin{longtable}{|c|cc|ccccc|}
\caption{Competition among various decay modes in the range 111$\leq$Z$\leq$118. $\#$ values are the estimated from Trends in Neighboring Nuclei (TNN) in NUBASE2020 \cite{audi20201}.} \label{tab:competition-decay-modes}\\

\hline

\multicolumn{1}{|c}{Nucleus}&
\multicolumn{2}{|c}{$log_{10}T_{1/2}$(s)}&
\multicolumn{5}{|c|}{Branching ratio} \\
\cline{2-8}

\multicolumn{1}{|c}{}&
\multicolumn{1}{|c}{Exp.}&
\multicolumn{1}{c}{Th.}&
\multicolumn{1}{|c}{$\alpha$}&
\multicolumn{1}{c}{Cluster}&
\multicolumn{1}{c}{SF}&
\multicolumn{1}{c}{$\beta^{-}$}&
\multicolumn{1}{c|}{$\beta^{+}$/EC}\\
\endfirsthead

\multicolumn{8}{c}%
{{\bfseries \tablename\ \thetable{} -- continued from previous page}} \\
\hline
\multicolumn{1}{|c}{}&
\multicolumn{1}{|c}{Exp.}&
\multicolumn{1}{c}{Th.}&
\multicolumn{1}{|c}{$\alpha$}&
\multicolumn{1}{c}{Cluster}&
\multicolumn{1}{c}{SF}&
\multicolumn{1}{c}{$\beta^{-}$}&
\multicolumn{1}{c|}{$\beta^{+}$/EC}\\

\hline
\endhead

\hline \multicolumn{8}{|r|}{{Continued on next page}} \\ \hline
\endfoot

\hline \hline
\endlastfoot                                                                                                                                   \hline
$^{280}$Rg&0.63$\pm$ 0.05        &-1.12&69.07$^{+1.13}_{-1.16}$  & 0.00&30.91$^{+1.13}_{-1.16}$  & 0.00 &0.01                     \\
$^{281}$Rg&1.23$^{+0.16}_{-0.08}$&-0.20& 6.67$^{+1.52}_{-1.26}$  & 0.00&93.19$^{+1.54}_{-1.27}$  & 0.07 &0.06$\pm$0.02   \\
$^{282}$Rg&2.11$\pm$ 0.18        &0.31 &60.80$^{+1.37}_{-1.39}$  & 0.00&38.98$^{+1.40}_{-1.42}$  & 0.00 &0.22$\pm$0.03  \\
$^{283}$Rg&2.08$^{\#}$           &0.50 &22.12$^{+2.14}_{-2.01}$  & 0.00&77.65$^{+2.19}_{-2.05}$  & 0.07$\pm$0.01 &0.16$\pm$0.05\\
$^{284}$Rg&1.78$^{\#}$           &1.74 &65.64$^{+9.78}_{-11.54}$& 0.00&31.03$^{+9.32}_{-11.25}$ & 0.00 &3.32$^{+0.46}_{-0.29}$  \\
$^{285}$Rg&1.48$^{\#}$           &2.27&66.89$^{+10.21}_{-12.59}$ & 0.00&30.50$^{+10.25}_{-12.83}$& 0.45$^{+0.18}_{-0.25}$ &2.15$^{+0.22}_{-0.49}$\\
$^{286}$Rg&1.00$^{\#}$           &3.00&69.39$^{+8.73}_{-10.49}$ & 0.00&6.69 $^{+2.88}_{-4.67}$  & 0.00         &23.92$^{+5.86}_{-5.83}$\\
$^{287}$Rg& -                    &3.02 &10.05$^{+1.16}_{-5.84}$  & 0.00&89.56$^{+12.45}_{-6.17}$  & 0.02$\pm$0.01 &0.37$^{+0.09}_{-0.03}$   \\
$^{288}$Rg&-                     &3.43&44.29$^{+18.91}_{-18.27}$ &0.00&39.30$^{+16.89}_{-19.72}$&0.39$^{+0.14}_{-0.13}$&16.01$^{+1.89}_{-1.58}$\\
$^{289}$Rg&-                     &1.26&0.04$\pm$0.02             &0.00&99.96$^{+0.06}_{-0.02}$   &0.00          &0.00\\
$^{290}$Rg&-                     &2.64&0.37$\pm$0.24   &0.00&98.07$^{+1.06}_{-0.51}$   &1.30$^{+0.08}_{-0.09}$ &0.26$\pm$0.18\\
$^{291}$Rg&-                     &1.76&0.01                      &0.00&99.98$^{+0.02}_{-0.01}$   &0.01          &0.00\\
$^{292}$Rg&-                     &2.66&0.08$\pm$0.05             &0.00&81.19$^{+0.83}_{-0.75}$   &18.73$^{+0.69}_{-0.70}$&0.00\\
$^{293}$Rg&-                     &1.45&0.00                      &0.00&99.75$\pm$0.02   &0.25$\pm$0.02 &0.00\\
$^{294}$Rg&-                     &1.92&0.00                      &0.00&88.42$^{+0.40}_{-0.39}$   &11.58$^{+0.40}_{-0.39}$&0.00\\
\hline
$^{278}$Cn&-2.70$^{\#}$           & -3.13&95.46$^{+0.38}_{-0.42}$ &0.00                     &4.53$^{+0.38}_{-0.42}$  &0.01&0.00\\
$^{279}$Cn&-4.22$^{\#}$           & -2.05&94.76$^{+0.46}_{-0.50}$ &0.00                     &5.20$^{+0.45}_{-0.50}$  &0.04&0.00\\
$^{280}$Cn&-2.30$^{\#}$           & -5.10&0.08$\pm$0.01           &0.00                     &99.92$\pm$0.01          &0.00&0.00\\
$^{282}$Cn&-3.04$^{+0.17}_{-0.09}$& -4.26&0.03                    &0.00                     &99.97                   &0.00&0.00\\
$^{284}$Cn&-1.01$^{+0.09}_{-0.06}$& -2.49&0.11                    &0.01                     &99.89                   &0.00&0.00\\
$^{285}$Cn&1.45$^{+0.12}_{-0.09}$&1.14&59.49$^{+15.60}_{-18.16}$&0.04$\pm$0.01&38.34$^{+14.86}_{-17.34}$&1.56$^{+0.65}_{-0.82}$&0.57$^{+0.08}_{-0.01}$\\
$^{287}$Cn& 1.48$^{\#}$&2.09&92.29$^{+2.71}_{-4.40}$&0.07$^{+0.02}_{-0.03}$&5.02$^{+2.02}_{-3.28}$&1.66$^{+0.74}_{-0.13}$&0.95$^{+0.07}_{-0.20}$\\
$^{288}$Cn& 1.00$^\#$             &-0.16&0.44          &0.00&99.52$\pm$0.44&0.04&0.00 \\
$^{289}$Cn&-                      &1.91 &49.52$^{+19.69}_{-19.95}$&0.01&50.40$^{+19.70}_{-19.93}$&0.05$^{+0.02}_{-0.03}$&0.01 \\
$^{290}$Cn&-                      &0.09 &0.23$\pm$0.03 &0.00&99.75$^{+0.31}_{-0.14}$&0.01&0.00 \\
$^{291}$Cn&-                      &3.20 &40.71$^{+22.08}_{-19.23}$&0.00&59.28$^{+22.08}_{-19.23}$&0.00&0.00 \\
$^{292}$Cn&-                      &0.63 &0.01          &0.00&99.98$^{+0.02}_{-0.01}$&0.00&0.00 \\
$^{293}$Cn&-                       &3.67 &4.44$^{+0.65}_{-0.27}$&0.00&95.56$^{+6.45}_{-2.74}$&0.00&0.00 \\
\hline
$^{285}$Nh&0.62$^{+0.15}_{-0.08}$&0.16&99.13$^{+0.04}_{-0.03}$&0.06&0.04&0.43$\pm$0.03&0.34$^{+0.07}_{-0.06}$\\
$^{286}$Nh&1.08$\pm$ 0.19        &0.43&99.42$\pm$0.08&0.08$^{+0.01}_{-0.02}$&0.00&0.00&0.50$^{+0.06}_{-0.07}$ \\
$^{287}$Nh&1.30$^{\#}$           &1.07&97.22$^{+0.05}_{-0.02}$&0.78$\pm$0.01&0.01&0.89$^{+0.14}_{-0.16}$&1.11$\pm$0.19 \\
$^{288}$Nh&1.30$^{\#}$           &0.99&98.88$^{+1.11}_{-6.84}$&0.21&0.00&0.00&0.90 \\
$^{289}$Nh&1.48$^{\#}$&2.69&74.40$^{+2.51}_{-7.22}$&8.65$^{+8.51}_{-2.92}$&0.12$^{+0.11}_{-0.05}$&3.86$^{+3.81}_{-1.64}$&12.97$^{+1.27}_{-2.61}$\\
$^{290}$Nh&0.90$\pm$ 0.42        &1.55&98.48$^{+0.46}_{-0.63}$&0.05$^{+0.02}_{-0.03}$&0.00&0.00&1.48$^{+0.44}_{-0.60}$ \\
$^{291}$Nh&-       &4.02&30.73$^{+3.70}_{-2.47}$&2.40$^{+1.19}_{-0.26}$&0.18$^{+0.10}_{-0.02}$&4.31$^{+2.60}_{-0.69}$&62.37$^{+0.19}_{-7.24}$\\
$^{292}$Nh&-                &3.31&66.38$^{+11.03}_{-12.62}$& 0.03$^{+0.01}_{-0.02}$  & 0.00 & 0.05$^{+0.02}_{-0.04}$ & 33.54$^{+10.99}_{-12.56}$ \\
$^{293}$Nh&-                     &4.34&99.80$^{+0.12}_{-0.30}$& 0.05$\pm$0.03 & 0.15$\pm$0.09 & 0.00 & 0.00 \\
$^{294}$Nh&-                     &4.28&45.78$^{+3.36}_{-1.92}$& 0.00  & 0.03 & 12.87$^{+2.83}_{-4.78}$& 41.32$^{+0.81}_{-0.67}$ \\
$^{295}$Nh&-                     &5.38&91.84$^{+1.07}_{-1.13}$& 0.01  & 8.15$^{+1.07}_{-1.22}$ & 0.00 & 0.00 \\
$^{296}$Nh&-                     &4.12&1.13$^{+0.15}_{-0.13}$ & 0.00  & 0.19$\pm$ 0.01 & 98.67$^{+0.18}_{-0.14}$& 0.01 \\
$^{297}$Nh&-                     &4.57&1.53$^{+0.22}_{-0.20}$ & 0.00  & 57.27$^{+1.38}_{-1.37}$& 41.21$^{+1.15}_{-1.17}$& 0.00 \\
$^{298}$Nh&-                     &1.97&0.71$^{+0.11}_{-0.09}$ & 0.00  & 78.09$^{+0.38}_{-0.37}$& 21.20$\pm$ 0.27& 0.00 \\
\hline
$^{282}$Fl & -                        & -3.44 & 99.06$^{+0.42}_{-0.77}$ & 0.00 & 0.93$^{+0.41}_{-0.76}$   & 0.00   & 0.00\\
$^{283}$Fl & -                        & -1.80 & 99.68$^{+0.17}_{-0.37}$ & 0.00 & 0.01   & 0.27   & 0.03\\
$^{285}$Fl & -0.88$^{+0.10}_{-0.22}$  & -1.20 & 99.71$^{+0.03}_{-0.04}$ & 0.05 & 0.00   & 0.21$^{+0.04}_{-0.05}$   & 0.03\\
$^{286}$Fl & -0.92$^{+0.15}_{-0.07}$  & -0.74 & 97.77$^{+0.10}_{-0.12}$ & 0.26$\pm$ 0.02 & 1.26$\pm$ 0.02   & 0.71$\pm$ 0.12   & 0.00\\
$^{287}$Fl & -0.32$^{+0.13}_{-0.08}$  & -0.28 & 97.43$^{+0.01}_{-0.05}$ & 1.69$^{+0.14}_{-0.13}$ & 0.00   & 0.77$\pm$ 0.03  & 0.11\\
$^{288}$Fl & -0.18$^{+0.09}_{-0.07}$&-0.05&94.04$^{+0.25}_{-0.35}$&3.89$\pm$0.14&0.34$^{+0.02}_{-0.03}$&1.73$^{+0.14}_{-0.15}$ & 0.00\\
$^{289}$Fl &  0.28$^{+0.17}_{-0.09}$  & 0.28  & 95.15$^{+0.13}_{-0.03}$ & 3.76$^{+0.30}_{-0.28}$ & 0.00   & 0.93$\pm$ 0.05 & 0.15\\
$^{290}$Fl &  1.90$\pm$ 0.42          & 0.48  & 86.73$^{+1.29}_{-1.59}$ & 10.68$^{+0.56}_{-0.58}$& 0.02   & 2.57$^{+0.41}_{-0.49}$ & 0.00\\
$^{291}$Fl & 1.00$^{\#}$  & 0.92  & 96.05$^{+3.94}_{-9.19}$ & 2.35$^{+2.34}_{-0.52}$ & 0.00   & 1.40$^{+1.39}_{-0.35}$   & 0.20$^{+0.20}_{-0.45}$\\
$^{292}$Fl & -& 2.32&20.52$^{+2.42}_{-2.24}$& 4.13$^{+0.04}_{-0.65}$  & 0.04  & 75.31$^{+2.23}_{-1.14}$  & 0.00\\
$^{293}$Fl & -& 3.21&48.68$^{+2.44}_{-2.69}$& 2.50$^{+0.76}_{-0.04}$  & 0.00  & 45.59$^{+2.72}_{-2.97}$  & 3.23$\pm$ 0.36\\
$^{294}$Fl & -& 2.91&14.41$^{+2.06}_{-1.85}$& 0.21  $\pm$ 0.02 & 0.16  & 85.22$^{+2.09}_{-1.87}$  & 0.00\\
$^{295}$Fl & -& 3.97&55.14$^{+4.43}_{-4.52}$& 0.23$\pm$ 0.01  & 0.01  & 44.61$^{+4.44}_{-4.53}$  & 0.00\\
$^{296}$Fl & -& 3.42&14.91$^{+2.24}_{-2.00}$& 0.01  & 1.87  & 83.21$^{+2.23}_{-2.00}$  & 0.00\\
$^{297}$Fl & -& 4.94&77.40$^{+3.54}_{-3.97}$& 0.04  & 0.34$\pm$ 0.03  & 22.22$^{+3.50}_{-3.94}$  & 0.00\\
$^{298}$Fl & -& 3.34&1.37$^{+0.22}_{-0.19}$ & 0.00  & 86.34$^{+0.15}_{-0.19}$ & 12.29$\pm$ 0.38  & 0.00\\
$^{299}$Fl & -& 2.52&14.49$^{+9.80}_{-10.87}$& 0.00  & 85.51$^{+9.80}_{-10.87}$ & 0.00   & 0.00\\
\hline
$^{289}$Mc & -0.48$^{+0.17}_{-0.12}$& -0.36 & 85.27$^{+0.88}_{-0.79}$ & 13.86$^{+0.96}_{-0.91}$ & 0.00   & 0.63$\pm$ 0.03 & 0.24\\
$^{290}$Mc & -0.08$\pm$ 0.20        & -0.61 & 90.14$^{+0.91}_{-1.00}$ & 9.76$^{+0.90}_{-0.99}$  & 0.00   & 0.00 & 0.09\\
$^{291}$Mc & 0.00$^{\#}$            & -0.39 & 57.15$^{+2.16}_{-2.12}$ & 42.49$^{+2.19}_{-2.16}$ & 0.00   & 0.22$^{+0.06}_{-0.08}$ & 0.14\\
$^{292}$Mc & 0.70$^{\#}$            & -0.10 & 93.07$^{+6.88}_{-8.89}$ & 6.72$^{+6.67}_{-0.86}$  & 0.00   & 0.00 & 0.21$^{+0.20}_{-0.28}$\\
$^{293}$Mc & -          & 1.19  & 73.35$^{+1.09}_{-1.58}$ & 21.65$^{+0.07}_{-0.06}$ & 0.00   & 2.66$^{+0.33}_{-0.38}$ &2.34$^{+0.13}_{-0.14}$\\
$^{294}$Mc & -                      & 1.22  & 95.65$^{+1.86}_{-3.81}$ & 1.94$\pm$ 0.02  & 0.00   & 0.00 & 2.41$^{+0.19}_{-0.20}$\\
$^{295}$Mc & -& 1.32& 97.40$^{+1.07}_{-2.27}$ & 0.46  & 0.00  & 0.68$^{+0.09}_{-0.11}$ & 1.46$\pm$ 0.07\\
$^{296}$Mc & -& 1.49& 97.36$^{+0.19}_{-0.20}$ & 0.04  & 0.00  & 0.00 & 2.60$^{+0.19}_{-0.20}$\\
$^{297}$Mc & -& 1.55& 99.01$^{+0.25}_{-0.53}$ & 0.01  & 0.00  & 0.19$\pm$ 0.03 & 0.79$\pm$ 0.01\\
$^{298}$Mc & -& 1.70& 98.03$^{+0.66}_{-0.88}$ & 0.00  & 0.00  & 0.00 & 1.97$^{+0.12}_{-0.13}$\\
$^{302}$Mc & -&-1.26& 54.38$^{+2.55}_{-2.58}$ & 0.00  & 45.62$^{+2.55}_{-2.58}$ & 0.00 & 0.00\\

\hline
$^{278}$Lv&-                      &-5.91& 56.60$^{+1.84}_{-1.86}$&0.00 &45.40$^{+1.84}_{-1.86}$&0.00 & 0.00\\
$^{281}$Lv&-                      &-3.76& 99.97$^{+0.01}_{-0.02}$&0.01 &0.00 &0.02 & 0.00\\
$^{288}$Lv&-                      &-2.35& 93.69$^{+2.42}_{-3.87}$&6.24$\pm$ 0.04 &0.00 &0.07$\pm$ 0.01 & 0.00\\
$^{289}$Lv&-1.80$^{\#}$           &-1.87& 78.49$^{+1.49}_{-1.41}$&21.37$^{+1.50}_{-1.43}$&0.00 &0.12$\pm$ 0.01 & 0.01$\pm$ 0.01\\
$^{290}$Lv&-2.08$^{+0.20}_{-0.10}$&-2.01& 46.28$^{+1.68}_{-1.69}$&53.63$^{+1.69}_{-1.70}$&0.00 &0.08$\pm$ 0.01 & 0.00\\
$^{291}$Lv&-1.72$^{+0.63}_{-0.14}$&-2.18& 12.45$^{+0.82}_{-0.87}$&87.53$^{+0.82}_{-0.87}$&0.00 &0.02 & 0.00\\
$^{292}$Lv&-1.89$^{+0.26}_{-0.14}$&-2.31& 7.49$\pm$ 0.18 &92.49$\pm$ 0.19&0.00 &0.02 & 0.00\\
$^{293}$Lv&-1.24$^{+0.43}_{-0.13}$&-1.39& 24.82$^{+1.42}_{-1.48}$&75.09$^{+1.43}_{-1.49}$&0.00 &0.08 & 0.02\\
$^{294}$Lv&-                      &-1.21& 46.01$^{+0.03}_{-0.02}$&53.85$^{+0.04}_{-0.03}$&0.00 &0.13$^{+0.01}_{-0.02}$ & 0.00\\
$^{295}$Lv&-                      &-1.14& 92.78$^{+2.94}_{-4.83}$&7.17$\pm$ 0.02 &0.00 &0.04 & 0.01\\
$^{302}$Lv&-                      &-3.90& 98.34$^{+0.70}_{-1.21}$&0.00 &1.66$^{+0.34}_{-0.42}$ &0.00 & 0.00\\
$^{303}$Lv&-                      &-3.67& 39.25$^{+5.11}_{-5.78}$&0.00 &60.75$^{+1.68}_{-3.08}$&0.00 & 0.00\\

\hline
$^{289}$Ts& -                      &-3.37& 93.31$^{+2.36}_{-3.59}$&6.68$\pm$ 0.09 &0.00 & 0.01&0.00\\
$^{290}$Ts& -                      &-3.24& 89.00$^{+3.83}_{-5.65}$&11.00$\pm$ 0.12&0.00 & 0.00&0.00\\
$^{291}$Ts& -2.70$^{\#}$           &-2.96& 22.81$^{+1.42}_{-1.48}$&77.18$^{+1.42}_{-1.48}$&0.00 & 0.01&0.00\\
$^{292}$Ts& -2.00$^{\#}$           &-3.07& 29.81$^{+4.01}_{-2.32}$&70.19$^{+4.01}_{-2.32}$&0.00 & 0.00&0.00\\
$^{293}$Ts& -1.66$^{+0.17}_{-0.08}$&-3.03& 8.53$^{+0.58}_{-0.62}$ &91.47$^{+0.59}_{-0.62}$&0.00 & 0.00&0.00\\
$^{294}$Ts& -1.15$\pm$ 0.20        &-2.48& 20.49$^{+1.53}_{-1.46}$&79.50$^{+1.53}_{-1.46}$&0.00 & 0.00&0.00\\
$^{295}$Ts& -                      &-2.37& 32.12$\pm$ 0.16&67.87$^{+12.02}_{-10.22}$&0.00 & 0.01&0.00\\
$^{296}$Ts& -                      &-2.48& 90.22$^{+3.59}_{-5.46}$&9.78$\pm$ 0.10 &0.00 & 0.00&0.00\\
$^{297}$Ts& -                      &-2.52& 96.54$^{+1.30}_{-2.10}$&3.45$\pm$ 0.04 &0.00 & 0.00&0.00\\
$^{305}$Ts& -                      &-4.22& 12.14$^{+0.96}_{-0.90}$&0.00 &87.86$^{+7.37}_{-4.90}$& 0.00&0.00\\

\hline
$^{283}$Og & -                        & -4.46 & 94.08$^{+1.80}_{-2.57}$ & 5.92$\pm$ 0.13  & 0.00  & 0.01&	0.00\\
$^{287}$Og & -                        & -4.86 & 89.77$^{+3.21}_{-4.45}$ & 10.23$^{+0.32}_{-0.44}$ & 0.00  & 0.00&	0.00\\
$^{288}$Og & -                        & -4.61 & 72.42$^{+7.62}_{-9.38}$ & 27.58$^{+0.76}_{-0.94}$ & 0.00  & 0.00&	0.00\\
$^{289}$Og & -                        & -4.59 & 66.16$^{+8.78}_{-10.26}$ & 33.84$^{+0.88}_{-1.03}$ & 0.00  & 0.00&	0.00\\
$^{290}$Og & -                        & -4.65 & 57.16$^{+9.95}_{-10.77}$ & 42.84$^{+9.95}_{-10.77}$ & 0.00  & 0.00&	0.00\\
$^{291}$Og & -                        & -4.51 & 26.27$^{+0.93}_{-0.77}$ & 73.73$^{+9.27}_{-7.63}$ & 0.00  & 0.00&	0.00\\
$^{292}$Og & -                        & -4.64 & 12.27$\pm$ 0.15 & 87.73$^{+5.70}_{-4.14}$ & 0.00  & 0.00&	0.00\\
$^{293}$Og & -3.00$^{\#}$             & -4.69 & 7.62$^{+0.38}_{-0.26}$  & 92.38$^{+3.82}_{-2.66}$ & 0.00  & 0.00&	0.00\\
$^{294}$Og & -3.16$^{+0.71}_{-0.14}$  & -4.50 & 3.12$^{+0.22}_{-0.24}$  & 96.88$^{+0.22}_{-0.24}$ & 0.00  & 0.00&	0.00\\
$^{295}$Og & -0.17$\pm$ 0.47          & -4.09 & 3.61$^{+0.41}_{-0.20}$  & 96.39$^{+4.08}_{-1.99}$ & 0.00  & 0.00&	0.00\\
$^{296}$Og & - & -4.13 & 4.06$\pm$ 0.04  & 95.94$^{+2.36}_{-1.54}$ & 0.00 & 0.00&	0.00\\
$^{297}$Og & - & -3.77 & 42.49$^{+11.37}_{-10.83}$ & 57.50$^{+0.37}_{-0.36}$ & 0.00 & 0.00&	0.00\\
$^{298}$Og & - & -3.81 & 53.23$^{+10.91}_{-11.44}$ & 46.77$^{+0.41}_{-0.40}$ & 0.00 & 0.00&	0.00\\
$^{299}$Og & - & -3.20 & 82.76$^{+5.63}_{-7.77}$ & 17.24$\pm$ 0.22 & 0.00 & 0.00&	0.00\\
$^{300}$Og & - & -3.10 & 94.20$^{+2.08}_{-3.22}$ & 5.80$\pm$ 0.09  & 0.00 & 0.00&	0.00\\
$^{301}$Og & - & -3.05 & 98.78$^{+0.45}_{-0.72}$ & 1.22$\pm$ 0.02  & 0.00 & 0.00&	0.00\\
$^{304}$Og & - & -5.13 & 98.70$^{+0.51}_{-0.84}$ & 0.00  & 1.30$\pm$ 0.10 & 0.00&	0.00\\
$^{305}$Og & - & -5.01 & 49.38$^{+0.73}_{-1.76}$ & 0.00  & 50.62$^{+0.73}_{-1.75}$& 0.00&	0.00\\
$^{307}$Og & - & -3.42 & 29.58 & 0.00  & 70.42& 0.00&	0.00\\
  \hline
\end{longtable}
\end{center}
 }

In this paper, we follow Eqn. (\ref{tbeta}) to calculate half-lives for $\beta^-$-decay, whereas, Eqn. (\ref{tecfinal}) is used to calculate half-lives for electron capture (EC). These half-lives are plotted in Fig. \ref{fig:decaymodes} along with half-lives of $\alpha$-decay from NMHF formula (Eqn. (\ref{qf})) and half-lives of spontaneous fission from MBF formula (Eqn. (\ref{baoSF})) for the considered range of nuclei. Half-lives of cluster emission calculated by UDL formula (Eqn. (\ref{udl})) are also plotted for each isotope. It is important to point out here that each point of cluster decay half-life in Fig. \ref{fig:decaymodes} corresponds to the minima of each parabola in Fig. \ref{fig:clusters} for a given isotope. Therefore, the clusters participating in this contest are the clusters which are found most probable for a particular parent isotope. To analyze the half-lives of weak-decay microscopically, sensitivity to the unknown $Q$-energies is crucial \cite{Sarriguren2020jpg,Sarriguren2021plb,Sarriguren2022prc}.
For this, we have calculated the uncertainties in theoretical $Q$-values of $\beta^-$-decay and EC by using 113 and 78 experimental data in region Z$>$82 \cite{audi20202}, which are found to be $\pm$0.02 MeV and $\pm$0.20 MeV, respectively. We have estimated the weak-decay half-lives using these uncertainties, however, to make Figs. \ref{fig:clusters} and \ref{fig:decaymodes} unclutter, we have not displayed the error bars. \par

\begin{figure*}[!htbp]
\centering
\includegraphics[width=1.00\textwidth]{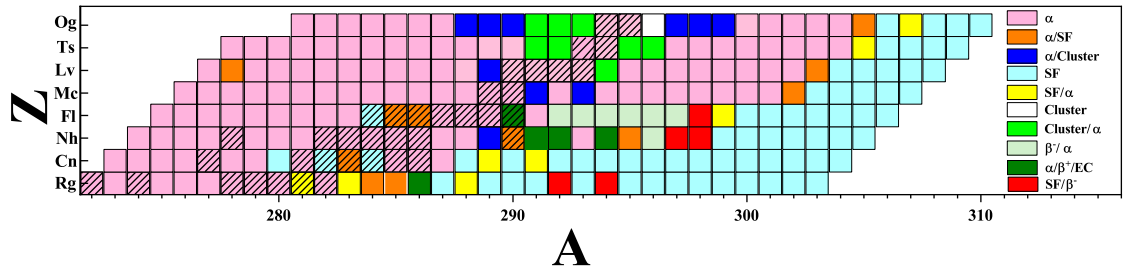}
\caption{(Colour online) Chart of considered nuclei with their probable decay modes. Shaded blocks show experimental decay modes taken from Ref. \cite{audi20201}.}
\label{fig:chart}
\end{figure*}

From Fig. \ref{fig:decaymodes} the competition among $\alpha$, cluster, SF, $\beta^{-}$ and EC decay modes is clearly evident. In principle, the decay mode which has the lowest half-life among others becomes most probable. In view of this, for the neutron rich isotopes, spontaneous fission is found to dominate while $\alpha$-decay prevails on the neutron deficient side for all the considered nuclei. In the middle region, however, the other decay modes begin to compete with $\alpha$ and SF. As an example, cluster decay (red line) is ascertained to be more probable than $\alpha$-decay for N$\sim$178 for Mc, Lv, Ts and Og isotopes. For Rg, Cn, Nh and Fl isotopes, weak-decay (pink and green lines) is observed to meet closely to the half-lives of other decay modes. To quantify this competition, we have calculated the branching ratios for respective decay modes as:

\begin{equation}\label{eq:branching}
  b  = \frac{T^{Th.}_{1/2}}{T^{\alpha/Cluster/SF/\beta/EC}_{1/2}}
\end{equation}
where, $T^{Th.}_{1/2}$ is the total half-life calculated by considering half-lives of all decay modes and the relation is given by:

\begin{equation}\label{eq:Total-hl}
  \frac{1}{T^{Th.}_{1/2}} = \frac{1}{T^{\alpha}_{1/2}}+\frac{1}{T^{Cluster}_{1/2}}+\frac{1}{T^{SF}_{1/2}}+\frac{1}{T^{\beta}_{1/2}}+\frac{1}{T^{EC}_{1/2}}
\end{equation}

where the superscripts refer to the half-lives of concerned decay modes. Respective branching ratios for considered decay modes
(from Eqn. (\ref{eq:branching})) are mentioned in percentage form with uncertainty in the Table \ref{tab:competition-decay-modes}. Experimental half-lives taken from NUBASE2020 \cite{audi20201} are mentioned along with the logarithm of total half-lives calculated by using Eqn. (\ref{eq:Total-hl}). These half-lives are found in a reasonable match which supports the approach of combining half-lives of various decay modes. From the close inspection of Table \ref{tab:competition-decay-modes}, chances of weak-decay are observed for Rg, Nh and Fl isotopes. Particularly, $^{294,296}$Fl is found with $\sim$ 90$\%$ probability of $\beta$-decay besides the $\gtrsim$50$\%$ probability of EC-decay in $^{291,294}$Nh. The chances of weak-decay, however, are found negligible for the nuclei with Z$>$114. Contrarily, cluster decay is found more probable in Z$>$114 which reaches up to 95$\%$ for the case of $^{294,295,296}$Og. At various other places this decay competes with $\alpha$-decay and becomes more significant. As a result,$^{83}$As, $^{84-86}$Se, $^{85-87}$Br, $^{86,88,89}$Kr clusters are found noteworthy to decay from $^{291}$Mc, $^{290-294}$Lv, $^{291-295}$Ts, $^{291-297}$Og nuclei, respectively. This important outcome has been summarized in Fig. \ref{fig:chart} as a periodic chart of the considered range. Most probable decay modes are mentioned by different colours in the diagram. Known decay modes as per NUBASE2020 \cite{audi20201} are pictured by shaded squares.

\section{Conclusions}
Theoretical calculations of half-lives of $\alpha$-decay, spontaneous fission, heavy-cluster decay and $\beta$-decay have been brought forward
by using NMHF, MBF, UDL and Fiset $\&$ Nix formulas, respectively. The $\alpha$-decay half-lives of synthesized SHN (Z$>$110) are successfully
reproduced by employing the NMHF formula which demonstrates the predictability of this formula in the superheavy region. NMHF formula has been used to calculate the favoured and unfavoured $\alpha$-decay half-lives of undetected SHN ranging from Rg to Og isotopes (161$\leq$N$\leq$192). In addition to the $\alpha$-decay half-lives, spontaneous fission half-lives (using MBF formula) are also calculated for various possible decay chains in the mentioned range, which are consequently utilized to predict various unknown $\alpha$-transitions and probable decay chains of SHN as well. \par

The cluster decay modes are also investigated in the mentioned range using UDL formula through the study and as a result $^{83}$As, $^{84-86}$Se,
$^{85-87}$Br, $^{86,88,89}$Kr cluster emissions from $^{291}$Mc, $^{290-294}$Lv, $^{291-295}$Ts, $^{291-297}$Og nuclei, respectively, are reported. Finally, a comparison among $\alpha$-decay, spontaneous fission, heavy-cluster decay and weak-decay modes is encased which establishes $\alpha$-decay and SF modes as the commanding mode of SHN, however, in few cases cluster decay mode is also found presiding over the $\alpha$ or SF decay modes. Additionally, chances of weak-decay modes are found equally probable in some nuclei and hence requires more detailed attention in this region of periodic chart too. Conclusively, we have done a comprehensive and combined study of various types of decay viz. $\alpha$-decay, SF-decay including rarely found weak-decay (in superheavy region) and the cluster decay in the above mentioned range with their accurate estimation of half-lives together with uncertainties and different branching ratios.
This theoretical study may provide a useful impetus for the detection of superheavy elements in the near future.

\section{Acknowledgement}
AJ and GS acknowledge the support provided by SERB (DST), Govt. of India under CRG/2019/001851 and SIR/2022/000566, respectively.

\newpage
\section{Appendix}
\begin{center}
\small
\setlength{\tabcolsep}{3pt}
\begin{table*}[!htbp]
\caption{$\alpha$-decay half-lives and prediction of decay chains of $^{281,301-305}$Og, $^{299,301,303}$Ts and $^{275}$Lv and their decay modes using NMHF formula. Q$_\alpha$-values (with uncertainties $\pm$0.04 MeV) are taken from WS4 mass model \cite{ws42014} and spin-parities are taken from Ref. \cite{moller2019}. Spontaneous Fission (SF) half-lives are calculated by MBF formula \cite{Saxena2021jpg}.}
\centering
\def\arraystretch{1.15}
\resizebox{1.00\textwidth}{!}{%
{\begin{tabular}{c|cccc|cc|c||c|cccc|cc|c}
 \hline
\multicolumn{1}{c}{Nucleus}&
\multicolumn{1}{|c}{Q$_\alpha$}&
\multicolumn{1}{c}{j$_{p}^{\pi}$}&
\multicolumn{1}{c}{j$_{d}^{\pi}$}&
\multicolumn{1}{c|}{$l$}&
\multicolumn{2}{c|}{log$_{10}$T$_{1/2}$(s)}&
\multicolumn{1}{c||}{Predicted}&
\multicolumn{1}{c}{Nucleus}&
\multicolumn{1}{|c}{Q$_\alpha$}&
\multicolumn{1}{c}{j$_{p}^{\pi}$}&
\multicolumn{1}{c}{j$_{d}^{\pi}$}&
\multicolumn{1}{c|}{$l$}&
  \multicolumn{2}{c|}{log$_{10}$T$_{1/2}$(s)}&
   \multicolumn{1}{c}{Predicted}\\
 \cline{6-7}\cline{14-15}
\multicolumn{1}{c}{}&
\multicolumn{1}{|c}{(MeV)}&
\multicolumn{1}{c}{}&
\multicolumn{1}{c}{}&
\multicolumn{1}{c|}{}&
\multicolumn{1}{c}{NMHF}&
\multicolumn{1}{c|}{MBF}&
\multicolumn{1}{c||}{Decay Mode}&
\multicolumn{1}{c}{}&
\multicolumn{1}{|c}{(MeV)}&
\multicolumn{1}{c}{}&
\multicolumn{1}{c}{}&
\multicolumn{1}{c|}{}&
\multicolumn{1}{c}{NMHF}&
\multicolumn{1}{c|}{MBF}&
\multicolumn{1}{c}{Decay Mode}\\
\multicolumn{1}{c}{}&
\multicolumn{1}{|c}{}&
\multicolumn{1}{c}{}&
\multicolumn{1}{c}{}&
\multicolumn{1}{c|}{}&
\multicolumn{1}{c}{($\alpha$)}&
\multicolumn{1}{c|}{(SF)}&
\multicolumn{1}{c||}{}& \multicolumn{1}{c}{}&
\multicolumn{1}{|c}{}&
\multicolumn{1}{c}{}&
\multicolumn{1}{c}{}&
\multicolumn{1}{c|}{}&
\multicolumn{1}{c}{($\alpha$)}&
\multicolumn{1}{c|}{(SF)}&
\multicolumn{1}{c}{}\\
 \hline
$^{305}$Og&12.91&5/2$^{+}$ &1/2$^{+}$ &2& -4.71$\pm$0.07          & -4.72 & $\alpha$  &$^{304}$Og&13.12&0$^{+}$ &0$^{+}$ &0& -5.13$\pm$0.07 &-3.25& $\alpha$ \\
$^{301}$Lv&11.58&1/2$^{+}$ &1/2$^{+}$ &0& -2.73$\pm$0.08          &  2.41 & $\alpha$  &$^{300}$Lv&10.92&0$^{+}$ &0$^{+}$ &0& -1.36$\pm$0.09 & 3.16& $\alpha$ \\
$^{297}$Fl& 8.35&1/2$^{+}$ &3/2$^{+}$ &2&  5.05$^{+0.22}_{-0.03}$ &  7.41 & $\alpha$  &$^{296}$Fl& 8.56&0$^{+}$ &0$^{+}$ &0&  4.25$\pm$0.12 & 5.15& $\alpha$ \\
$^{293}$Cn& 8.19&3/2$^{+}$ &1/2$^{+}$ &2&  5.02$^{+0.23}_{-0.04}$ &  3.69 & SF        &$^{292}$Cn& 8.28&0$^{+}$ &0$^{+}$ &0&  4.55$\pm$0.13 & 0.63& SF       \\
\hline
$^{303}$Og&12.61&7/2$^{+}$ &1/2$^{+}$ &4& -4.15$\pm$0.07          &  1.32 & $\alpha$  &$^{302}$Og&12.04&0$^{+}$ &0$^{+}$ &0& -3.21$\pm$0.08 &1.96& $\alpha$ \\
$^{299}$Lv&10.84&1/2$^{+}$ &3/2$^{+}$ &2& -1.05$^{+0.18}_{-0.01}$ &  7.23 & $\alpha$  &$^{298}$Lv&10.77&0$^{+}$ &0$^{+}$ &0& -1.06$\pm$0.09 &5.01& $\alpha$ \\
$^{295}$Fl& 8.61&3/2$^{+}$ &1/2$^{+}$ &2&  4.23$^{+0.22}_{-0.03}$ &  8.54 & $\alpha$  &$^{294}$Fl& 8.71&0$^{+}$ &0$^{+}$ &0&  3.75$\pm$0.12 &5.69& $\alpha$ \\
$^{291}$Cn& 8.59&1/2$^{+}$ &1/2$^{+}$ &0&  3.59$\pm$0.12          &  3.43 &SF/$\alpha$&$^{290}$Cn& 8.87&0$^{+}$ &0$^{+}$ &0&  2.72$\pm$0.12 &0.09& SF       \\
$^{287}$Ds& 7.77&1/2$^{+}$ &1/2$^{+}$ &0&  5.66$\pm$0.14          & -0.16 & SF        &-& -&-&- &-& - & -  & -      \\
\hline
$^{301}$Og&12.02&1/2$^{+}$ &3/2$^{+}$ &2& -3.04$^{+0.17}_{-0.02}$ & 6.06   & $\alpha$  &$^{281}$Og&13.76&13/2$^{-}$& 3/2$^{+}$  &5&-5.78$^{+0.66}_{-0.53}$& -1.02 &$\alpha$     \\
$^{297}$Lv&10.84&3/2$^{+}$ &1/2$^{+}$ &2& -1.07$^{+0.18}_{-0.01}$ & 8.49   & $\alpha$  &$^{277}$Lv&13.12& 3/2$^{+}$& 1/2$^{+}$  &2&-5.72$^{+0.16}_{-0.03}$& -1.04 &$\alpha$     \\
$^{293}$Fl& 8.78&1/2$^{+}$ &1/2$^{+}$ &0&  3.52$^{+0.12}_{-0.12}$ & 8.68   & $\alpha$  &$^{273}$Fl&12.97& 1/2$^{+}$&11/2$^{-}$  &5&-5.30$^{+0.66}_{-0.53}$& -3.51 &$\alpha$     \\
$^{289}$Cn& 9.05&1/2$^{+}$ &1/2$^{+}$ &0&  2.22$^{+0.11}_{-0.11}$ & 2.21   &SF/$\alpha$&$^{269}$Cn&12.57&11/2$^{-}$& 7/2$^{+}$  &3&-5.47$^{+0.28}_{-0.15}$& -4.71 &$\alpha$     \\
$^{285}$Ds& 7.80&1/2$^{+}$ &5/2$^{+}$ &2&  5.67$^{+0.23}_{-0.04}$ & 1.34 & SF          &$^{265}$Ds&12.33& 7/2$^{+}$&11/2$^{-}$  &3&-5.49$^{+0.28}_{-0.15}$& -2.70 &$\alpha$     \\
-         & -   &-         &-         &-&  -                      & -    &-            &$^{261}$Hs&10.96&11/2$^{-}$&9/2$^{-}$   &2&-3.51$^{+0.18}_{-0.01}$& -0.80 &$\alpha$     \\
-         & -   &-         &-         &-&  -                      & -    &-            &$^{257}$Sg& 9.71& 9/2$^{-}$&7/2$^{+}$   &1&-1.37$^{+0.12}_{-0.07}$& -0.44 &$\alpha$/SF  \\
\hline
$^{303}$Ts&12.78&1/2$^{-}$ &5/2$^{-}$ &2& -4.59$^{+0.16}_{-0.03}$ & 1.85 & $\alpha$  &$^{301}$Ts&11.61&1/2$^{-}$ &5/2$^{-}$&2&-2.44$^{+0.18}_{-0.02}$ & 7.11 & $\alpha$ \\
$^{299}$Mc& 9.59&5/2$^{-}$ &5/2$^{-}$ &0&  1.60$\pm$0.11          & 7.80 & $\alpha$  &$^{297}$Mc& 9.59&5/2$^{-}$ &5/2$^{-}$ &0&1.56$^{+0.10}_{-0.11}$& 9.58 & $\alpha$ \\
$^{295}$Nh& 8.15&7/2$^{-}$ &3/2$^{-}$ &2&  5.42$^{+0.23}_{-0.04}$ & 6.47 & $\alpha$  &$^{293}$Nh& 8.48&7/2$^{-}$ &3/2$^{-}$ &2&4.34$^{+0.22}_{-0.03}$ & 7.17 & $\alpha$ \\
$^{291}$Rg& 7.87&3/2$^{-}$ &1/2$^{+}$ &1&  5.71$^{+0.16}_{-0.11}$ & 1.76 & SF        &$^{289}$Rg& 8.30&3/2$^{-}$ &9/2$^{-}$ &4&4.68$^{+0.51}_{-0.25}$ & 1.26 & SF       \\
\hline
$^{299}$Ts&11.46& 1/2$^{-}$ & 5/2$^{-}$ &2& -2.16$^{+0.18}_{-0.01}$ &  9.02 & $\alpha$  &$^{275}$Lv&13.51& 1/2$^{+}$ &11/2$^{-}$&5&-5.82$^{+0.47}_{-0.34}$ & -5.24 & $\alpha$/SF \\
$^{295}$Mc& 9.72& 5/2$^{-}$ & 1/2$^{-}$ &2&  1.33$^{+0.20}_{-0.01}$ & 10.31 & $\alpha$  &$^{271}$Fl&13.40&11/2$^{-}$ & 9/2$^{+}$&1&-6.70$^{+0.56}_{-0.68}$ & -7.19 & $\alpha$/SF \\
$^{291}$Nh& 8.91& 1/2$^{-}$ &13/2$^{+}$ &7&  4.22$^{+1.28}_{-1.05}$ &  6.75 & $\alpha$  &$^{267}$Cn&13.05& 9/2$^{+}$ &11/2$^{-}$&1&-6.55$^{+0.60}_{-0.72}$ & -6.58 & $\alpha$/SF \\
$^{287}$Rg& 8.44&13/2$^{+}$ & 9/2$^{-}$ &3&  4.02$^{+0.34}_{-0.09}$ &  3.07 & SF        &$^{263}$Ds&12.34&11/2$^{-}$ & 9/2$^{-}$&2&-5.68$^{+0.38}_{-0.51}$ & -4.02 & $\alpha$ \\
-         & -   & -         &-          &-&  -    &  -    & -                           &$^{259}$Hs&10.79& 9/2$^{-}$ & 5/2$^{+}$&3&-3.08$^{+0.25}_{-0.09}$ & -2.79& $\alpha$/SF \\
-         & -   & -         &-          &-&  -    &  -    & -                           &$^{255}$Sg& 9.90& 5/2$^{+}$ & 5/2$^{+}$&0&-1.92$\pm$0.09          & -3.97 & SF      \\
\hline
\end{tabular}}
}
\label{Appendix}
\end{table*}
\end{center}
\end{document}